\documentclass[journal]{IEEEtran}

\usepackage{graphicx}
\usepackage[cmex10]{amsmath}
\usepackage{amsfonts}
\usepackage{amssymb}
\usepackage{mathrsfs}
\usepackage{bm}
\usepackage{bbold}
\usepackage{algorithmic}
\usepackage{array}
\usepackage{mdwmath}
\usepackage{mdwtab}
\usepackage[caption=false,font=footnotesize]{subfig}
\usepackage{fixltx2e}
\usepackage{stfloats}
\usepackage{footnote}
\usepackage{tabularx}
\usepackage{multirow}
\usepackage{color}
\usepackage{textcomp}
\usepackage{cite}
\usepackage{tikz}
\usetikzlibrary{mindmap,shadows,arrows,decorations.pathmorphing,backgrounds,positioning,fit,petri,calc}

\newcounter{MYtempeqncnt}

\newtheorem{definition}{Definition}
\newtheorem{proposition}{Proposition}
\newtheorem{theorem}{Theorem}
\newtheorem{lemma}{Lemma}

\newtheorem{assumption}{Assumption}
\newtheorem{example}{Example}

\title{Supply Function Equilibrium \\in Networked Electricity Markets}

\author{Yuanzhang~Xiao,~\IEEEmembership{Member,~IEEE,}
	Chaithanya~Bandi,
	and~Ermin~Wei,~\IEEEmembership{Member,~IEEE,}
	\thanks{Y. Xiao is with Hawaii Advanced Wireless Technologies Institute, University of Hawaii, Honolulu, HI 96816 (e-mail: xyz.xiao@gmail.com). The work was done when he was at Northwestern University.}%
	\thanks{C. Bandi is with Kellogg School of Management, Northwestern University, Evanston, IL 60208 USA (e-mail: c-bandi@kellogg.northwestern.edu).}%
	\thanks{E. Wei is with Department of Electrical Engineering and Computer Science, Northwestern University, Evanston, IL 60208 USA (e-mail: ermin.wei@northwestern.edu).}
}

\begin{document}
	\maketitle
	\begin{abstract}
		We study deregulated power markets with strategic power suppliers. In deregulated markets, each supplier submits its supply function (i.e., the amount of electricity it is willing to produce at various prices) to the independent system operator (ISO), who based on the submitted supply functions, dispatches the suppliers to clear the market with minimal total generation cost. If all suppliers reported their true marginal cost functions as supply functions, the market outcome would be efficient (i.e., the total generation cost is minimized). However, when suppliers are strategic and aim to maximize their own profits, the reported supply functions are not necessarily the true marginal cost functions, and the resulting market outcome may be inefficient. The efficiency loss depends crucially on the topology of the underlying transmission network. This paper provides an analytical upper bound of the efficiency loss due to strategic suppliers, and proves that the bound is tight under a large class of transmission networks (i.e., weakly cyclic networks). Our upper bound sheds light on how the efficiency loss depends on the transmission network topology (e.g., the degrees of nodes, the admittances and flow limits of transmission lines).
	\end{abstract}
	\begin{IEEEkeywords}
		Electricity markets, supply function bidding, supply function equilibrium, price of anarchy
	\end{IEEEkeywords}

	\section{Introduction}
	\label{sec:intro}
	A special feature of the power system is that supply and demand must be balanced at any time, because imbalance may cause serious consequences such as blackout \cite{Overbye-Book2012}. Therefore, due to the lack of large-scale energy storage, electricity markets become the major instrument in balancing supply and demand and maintaining the stability of power systems.
	
	Currently, most of the major electricity markets in United States and Europe are deregulated. In a deregulated market, the energy suppliers submit their {\it supply functions} to the independent system operator (ISO). The supply function specifies the amounts of electricity a supplier is willing to produce at different prices. The ISO considers the supply functions as the suppliers' marginal cost functions, and dispatches the suppliers such that the demand is met and the total generation cost is minimized. This procedure is called {\it economic dispatch}.
	
	If the suppliers submitted their true marginal cost functions as their supply functions, the economic dispatch would result in the socially optimal outcome that minimizes the total generation cost. However, the suppliers aim to maximize their own profits, and for this purpose, may choose supply functions that are different from their true marginal cost functions. In this case, the outcome of the economic dispatch is inefficient. The goal of this paper is to understand this inefficiency due to strategic behavior of energy suppliers.
	
	To analyze the efficiency loss in deregulated electricity markets, we first characterize the {\it supply function equilibrium (SFE)} of the market. We show that the supply profile (i.e., amounts of electricity produced by each supplier) at the equilibrium is unique. Following the literature, we define the efficiency loss as {\it price of anarchy (PoA)}, namely the ratio of the total generation cost at the SFE to the total cost at social optimum. By definition, the efficiency loss is larger when the PoA is larger. The main contribution of this paper is that we derive an analytical upper bound of the PoA with the following desirable properties:
	\begin{itemize}
		\item Our upper bound depends on, among other factors, the {\it topology of the underlying transmission networks} (e.g., the numbers of lines connected to each generator, the admittances and flow limits of transmission lines). Therefore, our results provide insights on how to optimize the transmission network in order to reduce the efficiency loss caused by strategic generators.
		\item Our bound is {\it provably ``tight''} under a large class of transmission networks, i.e., weakly cyclic networks (the networks in which one line belongs to at most one cycle).
	\end{itemize}
	
	The rest of this paper is organized as follows. We review the literature in Section~\ref{sec:related}. We describe our model of electricity markets and define the supply function equilibrium in Section~\ref{sec:model}. We analyze the equilibrium in Section~\ref{sec:analysis}. We provide numerical results in Section~\ref{sec:simulation}. Finally, Section~\ref{sec:conclusion} concludes the paper.
	
	\section{Related Works}\label{sec:related}

	Two types of models have been widely used to study strategic behavior in deregulated electricity markets. The first model is the Cournot competition model, where each generator submits the amount of electricity to produce (i.e., a scalar of quantity) \cite{Borenstein-RAND2000, Bose-CDC2014}. These works suggest that the network topology plays an important role in the efficiency loss. However, in the Cournot competition model, the generators act quite differently from the way they bid in reality. Hence, we want to analyze the efficiency loss under a model with a more realistic bidding format.
	
	The second model commonly used in the literature is the supply function equilibrium model, where each generator submits a supply function that specifies the amounts of electricity to produce at different prices\cite{KlempererMeyer-Econometrica1989, GreenNewbery-JPE1992, Green-JIE1996, Baldick-TPS2002, Baldick-JRegulatoryEcon2004, Baldick-JRegulatoryEcon2006, Wilson-OR2008, Anderson-MathProgram2013, JohariTsitsiklis-OR2011, XuLiLow-TPS2015, LiChenDahleh-TSG2015, XiaoBandiWei-Asilomar2015, XiaoBandiWei-Allerton2015, XiaoBandiWei-GlobalSIP2016, lin2019structural}. The SFE model is closer to the real bidding formats in electricity markets. Initial works on the SFE model allows the supply function to take an arbitrary form (with reasonable assumptions such as continuity) \cite{KlempererMeyer-Econometrica1989, GreenNewbery-JPE1992, Green-JIE1996, Baldick-TPS2002, Baldick-JRegulatoryEcon2004, Baldick-JRegulatoryEcon2006, Wilson-OR2008, Anderson-MathProgram2013}. Recent works parameterize the supply function with a scalar \cite{JohariTsitsiklis-OR2011, XuLiLow-TPS2015, LiChenDahleh-TSG2015, XiaoBandiWei-Asilomar2015, XiaoBandiWei-Allerton2015, XiaoBandiWei-GlobalSIP2016, lin2019structural}, resulting in more in-depth analysis of the efficiency loss at the equilibrium. However, most works using the SFE model do not study the impact of the transmission network topology on the efficiency loss \cite{JohariTsitsiklis-OR2011, XuLiLow-TPS2015, LiChenDahleh-TSG2015}. Their upper bounds of the PoA depend on the number of generators only \cite{JohariTsitsiklis-OR2011}, or on the number and the capacity limits of generators \cite{XuLiLow-TPS2015, LiChenDahleh-TSG2015}.
	
	The most closely related works are our prior works \cite{XiaoBandiWei-Asilomar2015, XiaoBandiWei-Allerton2015, XiaoBandiWei-GlobalSIP2016} and the work in \cite{lin2019structural}, which quantify the impact of the transmission network topology on the efficiency loss at SFE {\it to some extent}. Compared to our prior works \cite{XiaoBandiWei-Asilomar2015, XiaoBandiWei-Allerton2015, XiaoBandiWei-GlobalSIP2016}, this paper extends our analysis to general transmission network topology (as opposed to {\it radial} networks as in \cite{XiaoBandiWei-Asilomar2015, XiaoBandiWei-Allerton2015}).
	Compared to \cite{lin2019structural}, this paper gives a more explicit analytical expression of the upper bound of efficiency loss, and proves a tightness guarantee for our upper bound under a large class of networks (i.e., weakly cyclic networks). As we will show in Section~\ref{sec:analysis}, our bound recovers the bounds in existing works \cite{JohariTsitsiklis-OR2011, XuLiLow-TPS2015, XiaoBandiWei-Allerton2015} under the special cases considered therein.

    \section{System Model}\label{sec:model}
	
\begin{figure}
		\centering
		\centerline{\includegraphics[width=8.5cm]{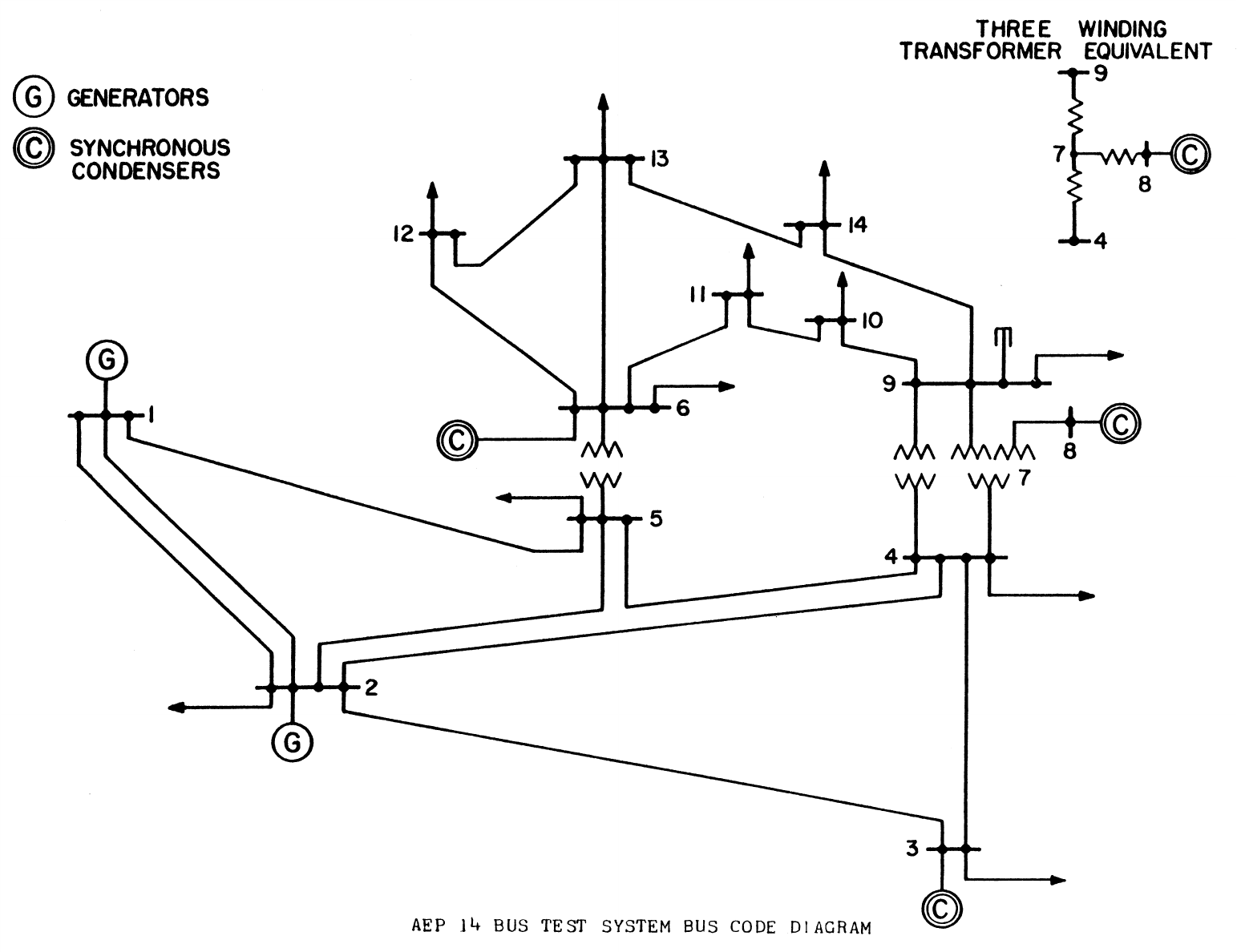}}
		\caption{Illustration of the IEEE 14-bus system, which will serve as the running example throughout the paper.}
		\label{fig:IEEE14bus}
	\end{figure}

	We model a power system as a graph $(\mathcal{N}, \mathcal{E})$, where each node in $\mathcal{N}$ is a bus\footnote{In this paper, we will use ``node'' and ``bus'' interchangeably.} with a generator or a load or both, and each edge in $\mathcal{E}$ is a transmission line connecting two buses. We assume that each generator is owned by a different supplier. Therefore, we can use ``generator'' and ``supplier'' interchangeably, and will use ``generator'' to emphasize the physical aspects and ``supplier'' to emphasize the strategic aspects. A representative power system, namely IEEE 14-bus system, is shown in Fig.~\ref{fig:IEEE14bus} \cite{IEEETestSystem}. Denote the set of buses that have a generator by $\mathcal{N}_g \subseteq \mathcal{N}$ (For the IEEE 14-bus system, we have $\mathcal{N}_g=\{1,2,3,6,8\}$). We assume that there are more than two generators (i.e., $\left|\mathcal{N}_g\right| > 2$).
	Since the majority of the load in the electricity market is inelastic \cite{MohsenianRad-TPS2016}, we assume that the load is inelastic, and denote the inelastic load profile by $\bm{d} = (d_1,\ldots,d_{|\mathcal{N}|})$.
	The total demand is then $D \triangleq \sum_{j \in \mathcal{N}} d_j$.
	
	Each generator $n \in \mathcal{N}_g$ has a cost of $c_n(s_n)$ in providing $s_n$ unit of electricity. We make the following standard assumption about cost functions.
	\begin{assumption}\label{assumption:cost-function}
		Each generator $n$'s cost function $c_n(s_n)$ is convex, strictly increasing, and differentiable in supply $s_n \in [0, +\infty)$.
	\end{assumption}
	
	Due to physical constraints, each generator $n$'s supply $s_n$ must be in a range $\left[\underline{s}_n, \bar{s}_n\right]$, namely 
	\begin{eqnarray}\label{eqn:capacity-constraint}
		\underline{s}_n \leq s_n \leq \bar{s}_n.
	\end{eqnarray}
	
	As in the literature \cite{XuLiLow-TPS2015, LiChenDahleh-TSG2015, XiaoBandiWei-Asilomar2015, XiaoBandiWei-Allerton2015, XiaoBandiWei-GlobalSIP2016, lin2019structural}, we assume that any generator is dispensable, in the sense that if it is removed from the market, the other generators have enough capacity to fulfill the demand. This assumption is formally stated below.
	\begin{assumption}\label{assumption:capacity-limit}
		With any generator $n$ removed from the market, the total capacity limit of the remaining generators is sufficient to fulfill the demand, namely $\sum_{m \in \mathcal{N}_g, m\neq n} \bar{s}_m > D$ for all $n \in \mathcal{N}_g$.
	\end{assumption}
	
	The supply profile $\bm{s} = (s_n)_{n \in \mathcal{N}_g}$ must satisfy physical constraints of the electrical network. First, in a power system, it is crucial to balance the supply and the demand at all time for the stability of the system \cite{Overbye-Book2012}. Hence, we need 
	\begin{eqnarray}\label{eqn:supply-equal-demand}
		\textstyle\sum_{n \in \mathcal{N}_g} s_n = D.
	\end{eqnarray}
	Second, the flow on each transmission line, which depends on the supply profile, cannot exceed the flow limit of the line. In economic dispatch, the ISO uses the linearized power flow model, where the flow on each line is the linear combination of injections from each node \cite{Overbye-Book2012}\cite{Bose-CDC2014}. Hence, the line flow constraints can be written as follows:
	\begin{eqnarray}\label{eqn:flow-constraint}
		-\bm{f} \leq \mathbf{A}_g \cdot \bm{s} + \mathbf{A}_\ell \cdot \bm{d} \leq \bm{f},
	\end{eqnarray}
	where $\bm{f} \in \mathbb{R}^{|\mathcal{E}|}$ is the vector of flow limits, $\mathbf{A}_g \in \mathbb{R}^{|\mathcal{E}| \times |\mathcal{N}_g|}$ and $\mathbf{A}_\ell \in \mathbb{R}^{|\mathcal{E}| \times |\mathcal{N}|}$ are shift-factor matrices. The shift-factor matrices $\mathbf{A}_g$ and $\mathbf{A}_\ell$ depend on the underlying transmission network topology (e.g., the degrees of nodes and the admittance of transmission lines). 
	
	\subsection{Benchmark - Social Optimum}
	If the ISO knew the true cost functions of each generator, it would solve for the optimal supply profile $\bm{s}^*$ to minimize the total generation cost subject to the constraints \eqref{eqn:capacity-constraint}--\eqref{eqn:flow-constraint}. We summarize this optimization problem as follows:
	\begin{eqnarray}\label{eqn:cost-minimization}
		& \displaystyle\min_{\bm{s}} & \textstyle\sum_{n \in \mathcal{N}_g} c_n(s_n) \\
		& s.t.                    & \textstyle\sum_{n \in \mathcal{N}_g} s_n = D, \nonumber \\
		&                          & \underline{s}_n \leq s_n \leq \bar{s}_n, \quad\forall n \in \mathcal{N}_g, \nonumber \\
		&                          & -\bm{f} \leq \mathbf{A}_g \cdot \bm{s} + \mathbf{A}_\ell \cdot \bm{d} \leq \bm{f}. \nonumber
	\end{eqnarray}
	
	To avoid triviality, we assume that the feasible set of supply profiles is non-empty and is not a singleton.
	\begin{assumption}\label{assumption:strictly-feasibility}
		There exists a strictly feasible supply profile $\bm{s}$ (i.e., a profile $\bm{s}$ that satisfies constraints \eqref{eqn:capacity-constraint}--\eqref{eqn:flow-constraint} strictly).
	\end{assumption}
	
	We write a solution to the optimization problem \eqref{eqn:cost-minimization} as $\bm{s}^* = (s_n^*)_{n \in \mathcal{N}_g}$, and call it the socially optimal supply profile.

	\subsection{Deregulated Markets and Supply Function Bidding}
	In practice, each supplier submits a supply function (i.e., a bid) to the ISO. A supply function is a mapping from the unit selling price of electricity to the amount of electricity produced by a supplier. In practice, the supply function is usually a step function. For analytical tractability, we assume that each supplier $n$ submits a parametrized supply function of the following form: \cite{JohariTsitsiklis-OR2011,XuLiLow-TPS2015,XiaoBandiWei-Allerton2015,lin2019structural}
	\begin{eqnarray}\label{eqn:supply-function}
		S_n(p, w_n) = D - \frac{w_n}{p},
	\end{eqnarray}
	where $w_n \in \mathbb{R}_+$ is supplier $n$'s strategic action, and $p \in \mathbb{R}_+$ is the unit price of electricity. To clear the market, (i.e., to find the price $p$ that satisfies the condition $\sum_{n\in\mathcal{N}_g} S_n(p, w_n) = D$), the ISO sets the price $p$ as follows:
	\begin{eqnarray}\label{eqn:market-clearing-price}
		p(\bm{w}) = \frac{\sum_{n \in \mathcal{N}_g} w_n}{\left(|\mathcal{N}_g|-1\right) D}.
	\end{eqnarray}
	where $\bm{w} = (w_n)_{n \in \mathcal{N}_g} \in \mathbb{R}_+^{\left|\mathcal{N}_g\right|}$ is the action profile.
	
	Each supplier $n$ aims to maximizes its profit $u_n\left( w_n, \bm{w}_{-n} \right)$, where $\bm{w}_{-n}$ is the action profile of all the suppliers other than $n$. Supplier $n$'s profit is the revenue minus cost as follows:
	\begin{eqnarray}\label{eqn:payoff-function}
		u_n\left( w_n, \bm{w}_{-n} \right) &=&
		\underbrace{p\left(w_n, \bm{w}_{-n}\right) \cdot S_n\left[ p\left(w_n, \bm{w}_{-n}\right), w_n \right]}_{\text{revenue}} \nonumber \\
		&-& \underbrace{c_n \left( S_n\left[p(w_n, \bm{w}_{-n}), w_n \right] \right)}_{\text{cost}}.
	\end{eqnarray}
	
	Now we formally define the supply function equilibrium.
	\begin{definition}
		An action profile $\bm{w}^{**}$ is a supply function equilibrium, if each supplier $n$'s action $w_n^{**}$ is a solution to the following profit maximizing problem:\footnote{We denote the $n$th column of a matrix $\mathbf{A}$ by $\left[\mathbf{A}\right]_{*n}$.} 
		\begin{eqnarray}\label{eqn:payoff-maximization}
			& \texttt{(SFE)}: & \nonumber \\
			&\displaystyle\max_{w_n} & u_n\left( w_n, \bm{w}_{-n}^{**} \right) \\
			& s.t.                                  & \underline{s}_n \leq S_n\left[ p\left(w_n, \bm{w}_{-n}^{**}\right), w_n \right] \leq \bar{s}_n, \forall n \in \mathcal{N}_g \nonumber \\
			&                                       & -\bm{f} \leq \nonumber \\
			&                                       & ~~~~~~~~\mathbf{A}_\ell \cdot \bm{d} + \left[\mathbf{A}_g\right]_{*n} \cdot S_n\left[ p(w_n, \bm{w}_{-n}^{**}), w_n \right] \nonumber \\
			&                                       & ~~~~~~~ + \sum_{\substack{m \in \mathcal{N}_g\\m \neq n}} \left[\mathbf{A}_g\right]_{*m} \cdot S_m\left[ p(w_n, \bm{w}_{-n}^{**}),  
			w_m^{**} \right] \nonumber \\
			&                                       &  ~~~~\leq \bm{f}. \nonumber
		\end{eqnarray}
	\end{definition}
	
	In a SFE, each supplier's action maximizes its own profit given the others' actions. Note that the set of feasible actions of each supplier depends on the others' actions. Therefore, the SFE is a generalized Nash equilibrium \cite{FacchineiKanzow-AnnalOR2010}.
	
	\section{Efficiency Loss at SFE}\label{sec:analysis}
	\subsection{Uniqueness of Equilibrium Supply Profile}
	Now we will show that the SFE exists and that there is a unique equilibrium supply profile at any SFE.
	
	\begin{proposition}\label{proposition:characterization-SFE}
		The SFE exists. At any SFE, the resulting equilibrium supply profile $\bm{s}^{**}$ is the unique solution to the modified cost minimization (MCM) problem
		\begin{eqnarray}\label{eqn:MCM}
			& \texttt{(MCM)}: & \nonumber \\
			\label{eqn:ModifiedSW-Objective}
			& \displaystyle\min_{\bm{s}} & \textstyle\sum_{n \in \mathcal{N}_g} \hat{c}_n(s_n) \\
			\label{eqn:ModifiedSW-MarketClear}
			& s.t. & \textstyle\sum_{n \in \mathcal{N}_g} s_n = D, \nonumber \\
			\label{eqn:ModifiedSW-GeneratorCapacity}
			&        & \underline{s}_n \leq s_n \leq \bar{s}_n, \forall n \in \mathcal{N}_g \nonumber \\
			\label{eqn:ModifiedSW-LineCapacity}
			&       &  -\bm{f} \leq \mathbf{A}_g \cdot \bm{s} + \mathbf{A}_\ell \cdot \bm{d} \leq \bm{f}, \nonumber
		\end{eqnarray}
		where 
		\begin{eqnarray}
			\hat{c}_n(s_n) 
			&=& 
			\left( 1 + 
			\frac{s_n}{\left(|\mathcal{N}_g|-2\right) D}
			\right) \cdot
			c_n(s_n) \nonumber \\
			&-& 
			\frac{1}{\left(|\mathcal{N}_g|-2\right) D} \cdot 
			\int_0^{s_n} c_n(x) dx. 
		\end{eqnarray}
	\end{proposition}
	\begin{IEEEproof} 
		See Appendix~\ref{proof:characterization-SFE}.
	\end{IEEEproof}

	Proposition~\ref{proposition:characterization-SFE} ensures the existence of SFE. In general, the generalized Nash equilibrium is not unique \cite{FacchineiKanzow-AnnalOR2010}. However, Proposition~\ref{proposition:characterization-SFE} indicates that although there may be multiple SFE $\bm{w}^{**}$, the resulting equilibrium supply profile $\bm{s}^{**}$ is always unique. This is important because the system efficiency is measured by the total generation cost, which depends on the supply profile, not on the action profile. As a result, the total generation cost at the equilibrium is unique.
	
	\subsection{Analysis of Efficiency Loss}
	We quantify the efficiency loss by PoA defined below:
	\begin{definition}\label{definition:PoA}
		PoA is the ratio of the total cost at SFE to the total cost at social optimum:
		$$
		\frac{\sum_{n \in \mathcal{N}_g} c_n(s_n^{**})}{\sum_{n \in \mathcal{N}_g} c_n(s_n^*)}.
		$$
	\end{definition}
	By definition, the PoA is never smaller than $1$. A larger PoA indicates a larger efficiency loss at the equilibrium. Note that the PoA is well defined because the equilibrium supply profile $\bm{s}^{**}$ is unique.
	
	\subsubsection{Preliminaries on Graph Theory}
	We derive an upper bound of the PoA that depends crucially on the topology of the transmission networks. Therefore, to better describe our bound, we need to introduce several useful concepts from graph theory. We will use the IEEE 14-bus system in Fig.~\ref{fig:IEEE14bus} as a running example to illustrate these concepts.
	
	\begin{definition}[Cycle]
		A cycle $C$ of a graph is a sequence of nodes $n_1, n_2, \ldots n_k, n_1$ ($k \geq 3$) that satisfies: 1) there is an edge between every consecutive nodes (i.e.,  $n_i$ and $n_{i+1}$ for $i=1,\ldots,k-1$) and between $n_k$ and $n_1$, and
		2) the nodes $n_1, \ldots, n_k$ are distinct.
	\end{definition}
	In short, a cycle is a group of distinct nodes that form a loop. 
	
	\begin{example}[Cycles]
		Take the IEEE 14-bus system in Fig.~\ref{fig:IEEE14bus} for example. The sequence $1,2,5,1$ forms a cycle. The sequence $2,1,5,2,4,3,2$ does not form a cycle (although all consecutive nodes are connected), because node 2 is visited twice in the loop.
	\end{example}
	
	Denote the set of node $n$'s neighbors by $\mathcal{N}(n)$. We then define a partition of $\mathcal{N}(n)$, denoted by $\mathcal{P}(n) \subset 2^{\mathcal{N}(n)}$.
	\begin{definition}[Partition of neighbors]\label{definition:partition}
		A partition $\mathcal{P}(n)$ is a set of singletons and duples of nodes that satisfy: 
		\begin{enumerate}
			\item the sets in $\mathcal{P}(n)$ are mutually exclusive, and the union of all sets in $\mathcal{P}(n)$ is $\mathcal{N}(n)$;
			\item any duple of nodes $\{i,j\} \in \mathcal{P}(n)$ are in a cycle with node $n$, namely $i,n,j$ or $j,n,i$ are in a cycle;
			\item any two singletones $\{i\}, \{j\} \in \mathcal{P}(n)$ are not in the same cycle.
		\end{enumerate}
	\end{definition}
	The partition $\mathcal{P}(n)$ divides node $n$'s neighbors into several subsets. Roughly speaking, we divide the neighbors by their affiliation to the cycles. Since node $n$ appears only once in a cycle, it has exactly two neighbors in the cycle. Therefore, each subset is either a duple of two nodes (in the case these two nodes are in the same cycle with node $n$), or a singleton (in the case this node is not in a cycle with either node $n$ or node $n$'s remaining neighbors). 
	
	Note that the partition is not unique, but any partition can be chosen for the purpose of deriving our upper bound on PoA. Once a partition $\mathcal{P}(n)$ has been chosen for each node $n \in \mathcal{N}_g$, we define a mapping $\mathcal{C}_n: \{i,j\} \mapsto C$ for each duple $\{i,j\} \in \mathcal{P}(n)$. The mapping indicates which cycle $C$ nodes $i,j,n$ belong to. In the case that nodes $i,j,n$ belong to multiple cycles, this mapping arbitrarily selects one of them. Again, this mapping is not unique, but any mapping can be chosen for our purpose. Hence, we will fix one mapping $\mathcal{C}_n$ for each node $n \in \mathcal{N}_g$ in the following.
	
\begin{figure}
		\centering
		\centerline{\includegraphics[width=8.5cm]{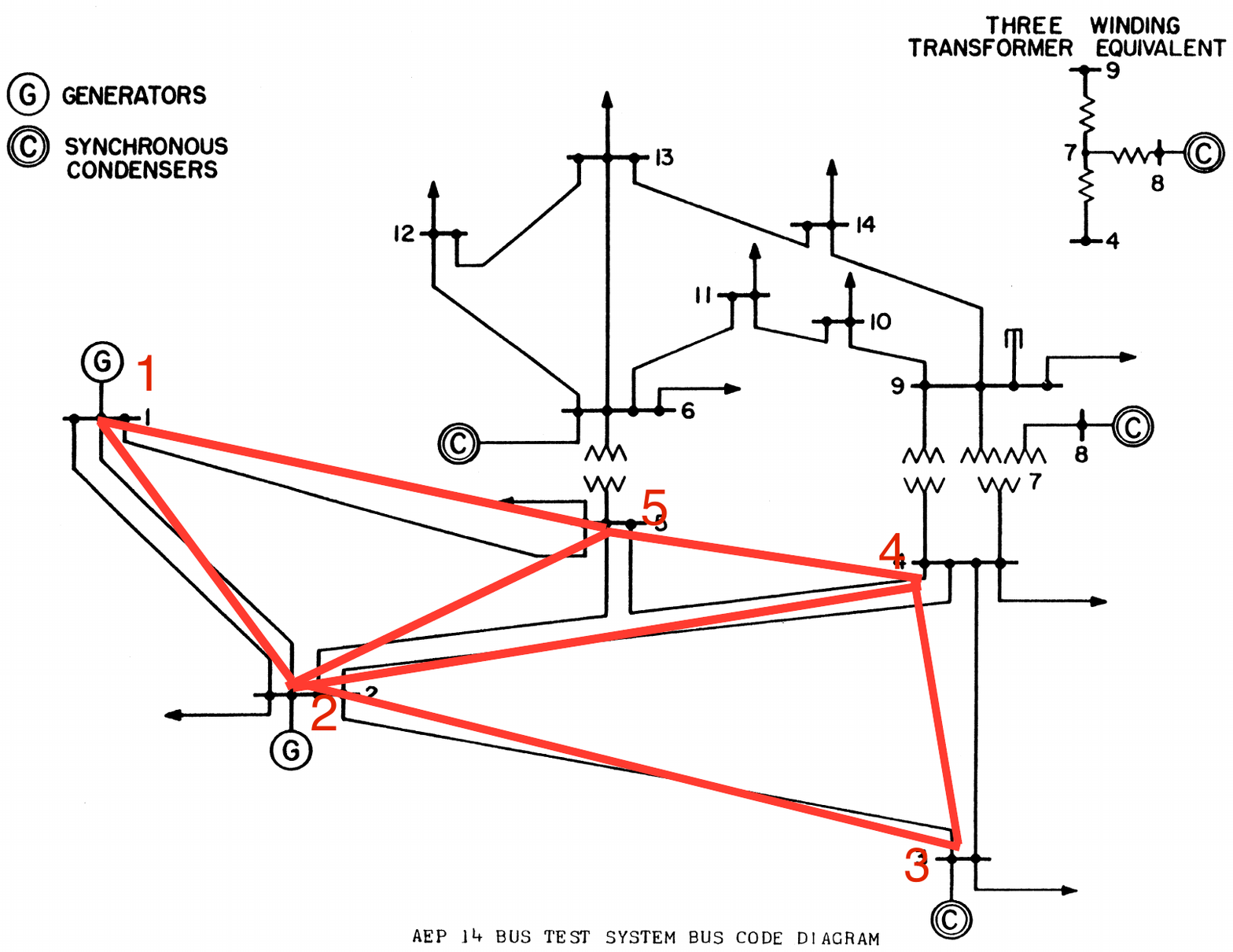}}
		\caption{Illustration of how we partition the neighbors and assign them to cycles for node $1$ and node $2$ of the IEEE 14-bus system. Node $1$'s neighbors are $\mathcal{N}(1) = \{2, 5\}$. The partition of $\mathcal{N}(1)$ is unique, namely $\mathcal{P}(1) = \{\{2,5\}\}$. The cycle assignment $\mathcal{C}_1$ is not unique: we can choose $\mathcal{C}_1(\{2,5\}) = 1,2,5,1$ or $1,2,4,5,1$. Node $2$'s neighbors are  $\mathcal{N}(2) = \{1,3,4,5\}$. The partition of $\mathcal{N}(2)$ is not unique. We could choose either $\mathcal{P}(2) = \{ \{1,5\}, \{3,4\} \}$ or $\mathcal{P}^\prime(2) = \{ \{1\}, \{3\}, \{4,5\} \}$. Under the partition $\mathcal{P}(2)$, the cycle assignment can be $\mathcal{C}_2(\{1,5\}) = 1,2,5,1$ and $\mathcal{C}_2(\{3,4\}) = 2,3,4,2$.}
		\label{fig:IEEE14bus-2}
	\end{figure}

	\begin{example}[Partition and Mapping to Cycles]
		Take the IEEE 14-bus system for example again. Node $1$ has two neighbors: nodes $2$ and $5$, which are in cycle $1,2,5,1$ with node $1$. Hence, the partition of node $1$'s neighbors is unique, namely $\mathcal{P}(1) = \{\{2,5\}\}$. Since nodes $1$, $2$ and $5$ belong to multiple cycles, the mapping $\mathcal{C}_1$ is not unique. In this case, we can choose $\mathcal{C}_1(\{2,5\})$ to be either $1,2,5,1$ or $1,2,4,5,1$. Node $2$ has four neighbors: nodes $1,3,4,5$. For node $2$, the partition of its neighbors is not unique. We could choose either $\mathcal{P}(2) = \{ \{1,5\}, \{3,4\} \}$ or $\mathcal{P}^\prime(2) = \{ \{1\}, \{3\}, \{4,5\} \}$. If we choose $\mathcal{P}(2)$, we can set the mapping $\mathcal{C}_2$ as $\mathcal{C}_2(\{1,5\}) = 1,2,5,1$ and $\mathcal{C}_2(\{3,4\}) = 2,3,4,2$. See Fig.~\ref{fig:IEEE14bus-2} for illustration.
	\end{example}
	
	All the above definitions of cycles, partitions, and mappings to cycles are preparations for the next crucial definition of \emph{effective flow limit}. For each supplier $n \in \mathcal{N}_g$, we define an effective flow limit $\hat{f}_{nm}$ for each outgoing link $m$ from supplier $n$. 
	
	\begin{definition}[Effective Flow Limit]
		For each supplier $n$, fix a partition $\mathcal{P}(n)$ of its neighbors and the mapping $\mathcal{C}_n$. Then the effective flow limit from $n$ to its neighbor $m \in \mathcal{N}(n)$ is defined as follows:
		\begin{itemize}
			\item If $m$ is a singleton in the partition (i.e., $\{m\} \in \mathcal{P}(n)$), the effective flow limit is the same as the conventional flow limit, namely $\hat{f}_{nm} = f_{nm}$;
			\item If $m$ is in a duple in the partition (i.e., $\{m,i\} \in \mathcal{P}(n)$), the effective flow limit is as follows:
			\begin{eqnarray}
				\hat{f}_{nm} = \min\left\{ f_{nm} , 
				\left( \sum_{\substack{j,k \in \mathcal{C}_n(\{m,i\}) \\ \{j,k\} \neq \{n,m\}}} \frac{f_{jk}}{B_{jk}}     
				\right)
				\cdot B_{nm} \right\},
			\end{eqnarray}
			where $B_{nm}$ is the admittance of the transmission line between bus $n$ and bus $m$.
		\end{itemize}
	\end{definition}
	The effective flow limit is the minimum between the conventional flow limit and another term that depends on the flow limits and admittances of the other lines in the cycle. The latter term is the flow from $n$ to $m$ when the other lines in the cycle reach flow limits. The flow from $n$ to $m$ cannot exceed this term, even if the flow limit $f_{nm}$ allows.
	
	Note that the effective flow limit depends on the direction of the flow, namely $\hat{f}_{nm} \neq \hat{f}_{mn}$ in general. This is because the partition of the neighbors and the associated cycles are different for different nodes $n$ and $m$. For example, for the IEEE 14-bus system in Fig.~\ref{fig:IEEE14bus}, we have
	$$
	\hat{f}_{12} = \min \left\{ f_{12} , \left( \frac{f_{25}}{B_{25}} + \frac{f_{51}}{B_{51}} \right) \cdot B_{12} \right\}
	$$
	under the partition $\mathcal{P}(1) = \{\{2,5\}\}$, and
	$$
	\hat{f}_{21} = f_{21} = f_{12}
	$$
	under the partition $\mathcal{P}^\prime(2) = \{ \{1\}, \{3\}, \{4,5\} \}$.
	
	\subsubsection{Analytical Upper Bound}
	With the effective flow limit defined, we are ready to provide an analytical upper bound of the PoA.
	\begin{theorem}\label{theorem:PoA}
		The PoA is upper bounded by 
		$$
		1 + 
		\max_{n \in \mathcal{N}_g} 
		\frac{ \min\left\{ \bar{s}_n, 
			D - \displaystyle\sum_{\substack{m \in \mathcal{N}_g \\ m \neq n}} \underline{s}_m, 
			d_n + \displaystyle\sum_{m \in \mathcal{N}(n)} \hat{f}_{nm} \right\} 
		}
		{ \left(|\mathcal{N}_g| - 2\right) D }.
		$$ 
	\end{theorem}
	\begin{IEEEproof}
		See Appendix~\ref{proof:PoA}.
	\end{IEEEproof}

	The upper bound in Theorem~\ref{theorem:PoA} gives us insights on the key factors that influence the efficiency loss. First, the upper bound is higher if one supplier has a significantly higher capacity limit than the others. In this case, this supplier may have larger market power, especially when the total capacity from the other suppliers are barely enough to fulfill the demand. Second, the upper bound is higher if one generator has higher local demand and higher effective flow limits of its outgoing links. In this case, this generator has advantage over the other generators in fulfilling its local demand (because incoming flow limits constrain the import of electricity from the other generators), and can more easily export its electricity generation to other nodes due to higher outgoing effective flow limits. Therefore, this generator has more influence on the market outcome.
	
	Our upper bound in Theorem~\ref{theorem:PoA} recovers the bounds in prior work as special cases. When all the generators have the same capacity limits of $\underline{s}_n = 0$ and $\bar{s}_n = D$ for all $n \in \mathcal{N}_g$, and when we ignore the network topology (i.e., ignore the term $d_n + \sum_{m \in \mathcal{N}(n)} \hat{f}_{nm}$), the bound reduces to the one in \cite{JohariTsitsiklis-OR2011}. If we allow generators to have different capacity limits $\bar{s}_n$, the bound reduces to the one in \cite{XuLiLow-TPS2015}. If we ignore the cycles in the transmission network, we have $\hat{f}_{nm} = f_{nm}$ for all $n$ and $m$, and hence recover the bound in our prior work \cite{XiaoBandiWei-Allerton2015}.

	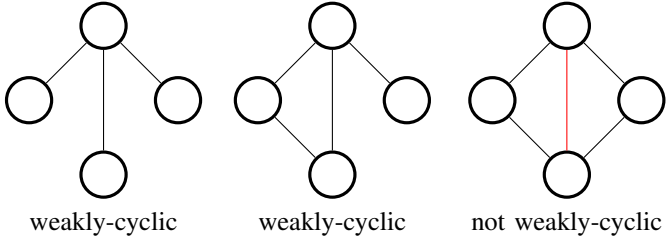
\begin{figure}\label{fig:weakly-cyclic}
		\begin{tikzpicture}[
			scale=0.33,
			root/.style={
				circle,
				minimum size=6mm,
				very thick,
				draw=black, 
				fill=white, 
				font=\large
			},
			branch1/.style={
				circle,
				minimum size=6mm,
				very thick,
				draw=red!50, 
				fill=red!20, 
				font=\large
			},
			branch2/.style={
				circle,
				minimum size=6mm,
				very thick,
				draw=blue!50, 
				fill=blue!20, 
				font=\large
			},
			branch3/.style={
				circle,
				minimum size=6mm,
				very thick,
				draw=black!50, 
				fill=black!20, 
				font=\large
			},
			output/.style={
				rectangle,minimum size=6mm,rounded corners=3mm,
				very thick,draw=black,
				font=\large},
			]

			\node (node1) at (0, 0) [root] {};
			\node (node2) at (-3, -3) [root] {};
			\node (node3) at (0, -6) [root] {};
			\node (node4) at (3, -3) [root] {};
			
			\draw [-] (node1) -- (node2);
			\draw [-] (node1) -- (node3);
			\draw [-] (node1) -- (node4);
			
			\node at (0, -8) {weakly-cyclic};
			
		\end{tikzpicture}
		\hspace{0.02\linewidth}
		\begin{tikzpicture}[
			scale=0.33,
			root/.style={
				circle,
				minimum size=6mm,
				very thick,
				draw=black, 
				fill=white, 
				font=\large
			},
			branch1/.style={
				circle,
				minimum size=6mm,
				very thick,
				draw=red!50, 
				fill=red!20, 
				font=\large
			},
			branch2/.style={
				circle,
				minimum size=6mm,
				very thick,
				draw=blue!50, 
				fill=blue!20, 
				font=\large
			},
			branch3/.style={
				circle,
				minimum size=6mm,
				very thick,
				draw=black!50, 
				fill=black!20, 
				font=\large
			},
			output/.style={
				rectangle,minimum size=6mm,rounded corners=3mm,
				very thick,draw=black,
				font=\large},
			]

			\node (node1) at (0, 0) [root] {};
			\node (node2) at (-3, -3) [root] {};
			\node (node3) at (0, -6) [root] {};
			\node (node4) at (3, -3) [root] {};
			
			\draw [-] (node1) -- (node2);
			\draw [-] (node1) -- (node3);
			\draw [-] (node1) -- (node4);
			\draw [-] (node2) -- (node3);
			
			\node at (0, -8) {weakly-cyclic};
			
		\end{tikzpicture}
		\hspace{0.02\linewidth}
		\begin{tikzpicture}[
			scale=0.33,
			root/.style={
				circle,
				minimum size=6mm,
				very thick,
				draw=black, 
				fill=white, 
				font=\large
			},
			branch1/.style={
				circle,
				minimum size=6mm,
				very thick,
				draw=red!50, 
				fill=red!20, 
				font=\large
			},
			branch2/.style={
				circle,
				minimum size=6mm,
				very thick,
				draw=blue!50, 
				fill=blue!20, 
				font=\large
			},
			branch3/.style={
				circle,
				minimum size=6mm,
				very thick,
				draw=black!50, 
				fill=black!20, 
				font=\large
			},
			output/.style={
				rectangle,minimum size=6mm,rounded corners=3mm,
				very thick,draw=black,
				font=\large},
			]

			\node (node1) at (0, 0) [root] {};
			\node (node2) at (-3, -3) [root] {};
			\node (node3) at (0, -6) [root] {};
			\node (node4) at (3, -3) [root] {};
			
			\draw [-] (node1) -- (node2);
			\draw [-,red] (node1) -- (node3);
			\draw [-] (node1) -- (node4);
			\draw [-] (node2) -- (node3);
			\draw [-] (node3) -- (node4);
			
			\node at (0, -8) {not weakly-cyclic};
			
		\end{tikzpicture}

		\caption{Illustration of weakly-cyclic networks. The first network is weakly-cyclic, because it is a tree and has no cycle. The second one is weak-cyclic, because no edge is shared by two or more cycles. The third one is not weakly-cycle, because the edge in red is shared by two cycles.}
		
	\end{figure}

	\subsubsection{Tightness}
	Our upper bound in Theorem~\ref{theorem:PoA} is tight under a large class of networks, namely weakly-cyclic networks. We formally defined weakly-cyclic networks below (see Fig.~2 for an illustration).
	
	\begin{definition}[Weakly-Cyclic Networks]
		A weakly-cyclic network is a graph in which no edge belongs to more than one cycle.
	\end{definition}

	The tightness result is described in the following theorem.
	
	\begin{theorem}\label{theorem:tightness}
		For any weakly-cyclic network, our upper bound is tight in the following sense: Given any small number $\varepsilon>0$, there are instances of demand profiles $\bm{d}$, sets of generators $\mathcal{N}_g$, cost functions $\{c_n\}_{n\in\mathcal{N}_g}$, generator capacity limits $\{\underline{s}_n, \bar{s}_n\}_{n\in\mathcal{N}_g}$, flow limits $\{f_{nm}\}_{nm \in \mathcal{E}}$, and admittances $\{B_{nm}\}_{nm \in \mathcal{E}}$, under which
		\begin{itemize}
			\item the upper bound of PoA is equal to
			$$
			1 + 
			\max_{n \in \mathcal{N}_g} 
			\frac{
				d_n + \textstyle\sum_{m \in \mathcal{N}(n)} \hat{f}_{nm}
			}
			{ \left(|\mathcal{N}_g| - 2\right) D }.
			$$ 
			\item the difference between the actual PoA and the upper bound is within $\varepsilon$.
		\end{itemize}
	\end{theorem}
	\begin{IEEEproof}
		See Appendix~\ref{proof:tightness}
	\end{IEEEproof}
	
	Our tightness result has several implications. The obvious implication is that for any weakly-cyclic network, one cannot improve our bound ``uniformly'' on all instances of demand profiles, sets of generators, cost functions, generator capacity limits, flow limits and admittances. In other words, there may exist bounds that are tighter for some instances, but not for all instances. The second implication is that the network topology matters. Recall that our bound in Theorem~\ref{theorem:PoA} is the minimum of three components. The first two components depend on the generator capacity limits, while the third component depends on the network topology (i.e., flow limits and admittances). Theorem~\ref{theorem:tightness} shows that for the instances where our bound is tight, our bound is equal to the third component. In other words, our bound is tight because we take into account the network topology.
	
	Finally, we note that a class of practical weakly-cyclic power networks are distribution networks (which are usually radial and contains no cycle). Today's distribution networks are evolving towards incorporating a large number of distributed energy resources (i.e., generators). Our result provides a tight prediction of the efficiency of electricity markets in future distribution networks.
	
	\begin{table}[!t]
		\renewcommand{\arraystretch}{1.2}
		\caption{Tightness of Our Bounds in The IEEE 1888-Bus Case}
		\label{table:simulation}
		\centering
		\begin{tabular}{c | c | c | c}
			\hline
			\% of congested lines & PoA & Our Bound & Existing Bound\\
			\hline
			0               & 1.0075 & 1.0521 & 1.0523 \\
			5               & 1.0075 & 1.0323 & 1.0523 \\
			10               & 1.0075 & 1.0243 & 1.0523 \\
			20               & 1.0079 & 1.0198 & 1.0523 \\
			30               & 1.0083 & 1.0145 & 1.0523 \\
			40               & 1.0085 & 1.0132 & 1.0523 \\
			60               & 1.0086 & 1.0095 & 1.0523 \\
			80               & 1.0088 & 1.0093 & 1.0523 \\
			90               & 1.0089 & 1.0091 & 1.0523 \\
			\hline
		\end{tabular}
	\end{table}

	\section{Simulation}\label{sec:simulation}
	
	We compare our bound with existing bounds and the true PoA under realistic test cases. We choose the IEEE 1888-bus test case (see Appendex~D.3 of MATPOWER User's Manual \cite{MATPOWER}), because this is a test case with a large number of buses and with line flow limits. We compare with the existing bound in \cite{XuLiLow-TPS2015}, which does not consider the network topology. Note that we cannot compare with the existing bound in \cite{JohariTsitsiklis-OR2011}, because their bound is derived based on the assumption that the generators have equal capacity limits, which is violated in the test cases.
	
	Table~\ref{table:simulation} summarizes the numerical results as we scale the flow limits so that the percentage of congested lines (at the equilibrium) varies. We can see that our bound is tighter than the network-independent bound in \cite{XuLiLow-TPS2015} in all the cases, and is much tighter in the cases with more congested lines. This is reasonable because if there are few congested lines, the network topology is not important and our bound cannot improve much on network-independent bounds. 
	

	\section{Conclusion}\label{sec:conclusion}
	We analyzed the efficiency loss in decentralized electricity markets. The distinct feature of our work is our consideration of the topology of the mesh transmission network. We show that there exists a unique equilibrium supply profile, and gave an analytical upper bound of the efficiency loss at the equilibrium. Our upper bound suggests that to reduce the efficiency loss, we should evenly distribute the generation capacity and outgoing effective flow limits among the generators. 
	
	\appendices
	
	\section{Proof of Proposition~\ref{proposition:characterization-SFE}}\label{proof:characterization-SFE}

	\begin{figure*}[!t]
		\normalsize
		\setcounter{MYtempeqncnt}{\value{equation}}
		\setcounter{equation}{\value{equation}}
		\begin{eqnarray}\label{eqn:SFE-KKT-Derivative-payoff}
			\frac{\partial u_n(w_n, \bm{w}_{-n})}{\partial w_n} -
			\left[ \underline{s}_n + \left( \left|\mathcal{N}_g\right| - 2 \right) D \right] \underline{\mu}_n^{\texttt{SFE}}  +
			\left[ \underline{s}_n + \left( \left|\mathcal{N}_g\right| - 2 \right) D \right] \bar{\mu}_n^{\texttt{SFE}} + 
			\left(\bm{\underline{\lambda}}^{\texttt{SFE}}\right)^T (\bm{b}_n+\bm{f}) - 
			\left(\bm{\bar{\lambda}}^{\texttt{SFE}}\right)^T (\bm{b}_n-\bm{f}) 
			= 0,
		\end{eqnarray}
		\begin{eqnarray}\label{eqn:SFE-KKT-Derivative-generation-capacity1}
			0 \leq 
			\underline{\mu}_n^{\texttt{SFE}} \perp 
			\left[ \underline{s}_n + \left( \left|\mathcal{N}_g\right| - 2 \right) D \right] w_n + 
			\left( \underline{s}_n - D \right) \textstyle\sum_{m\in\mathcal{N}_g, m \neq n} w_m 
			\leq 0, 
		\end{eqnarray}
		\begin{eqnarray}\label{eqn:SFE-KKT-Derivative-generation-capacity2}
			0 \leq \bar{\mu}_n^{\texttt{SFE}} \perp
			\left[ \bar{s}_n + \left( \left|\mathcal{N}_g\right| - 2 \right) D \right] w_n + 
			\left( \bar{s}_n - D \right) \textstyle\sum_{m\in\mathcal{N}_g, m \neq n} w_m 
			\geq 0,
		\end{eqnarray}
		\begin{eqnarray}\label{eqn:SFE-KKT-Derivative-line-capacity1}
			\bm{0} \leq 
			\bm{\underline{\lambda}}^{\texttt{SFE}} \perp 
			\textstyle\sum_{n\in\mathcal{N}_g} \left[ (\bm{b}_n+\bm{f}) \cdot w_n \right] 
			\geq \bm{0},
		\end{eqnarray}
		\begin{eqnarray}\label{eqn:SFE-KKT-Derivative-line-capacity2}
			\bm{0} \leq 
			\bm{\bar{\lambda}}^{\texttt{SFE}} \perp 
			\textstyle\sum_{n\in\mathcal{N}_g} \left[ (\bm{b}_n-\bm{f}) \cdot w_n \right] 
			\leq \bm{0} 
		\end{eqnarray}
		\setcounter{equation}{\value{equation}}
		\hrulefill
		\vspace*{4pt}
	\end{figure*}
	
	\subsection{Existence}
	We first prove the existence of SFE. Before the proof, we rewrite suppliers' payoff maximization problem \eqref{eqn:payoff-maximization}. Based on the supply function \eqref{eqn:supply-function} and the market clearing price \eqref{eqn:market-clearing-price}, we can calculate supplier $n$'s supply as
	\begin{eqnarray}\label{eqn:supply-as-function-of-bids}
		S_n\left[ p(w_n, \bm{w}_{-n}), w_n \right] = 
		D - \frac{w_n}{\sum_{m\in\mathcal{N}_g} w_m} \left( \left|\mathcal{N}_g\right| - 1 \right) D.
	\end{eqnarray}
	By substituting $S_n\left[ p(w_n, \bm{w}_{-n}), w_n \right]$ with its expression in \eqref{eqn:supply-as-function-of-bids} and separating supplier $n$'s action $w_n$ from other suppliers' bids $\bm{w}_{-n}$, supplier $n$'s payoff maximization problem in \eqref{eqn:payoff-maximization} can be rewritten as
	\begin{eqnarray}\label{eqn:SFE-reformulation}
		& \displaystyle\max_{w_n} & u_n(w_n, \bm{w}_{-n}) \\
		\label{eqn:SFE_GeneratorCapacity1}
		& s.t. & \left[ \underline{s}_n + \left( \left|\mathcal{N}_g\right| - 2 \right) D \right] w_n + \left( \underline{s}_n - D \right) \sum_{m \neq n} w_m \leq 0, \nonumber \\
		\label{eqn:SFE_GeneratorCapacity2}
		&        & \left[ \bar{s}_n + \left( \left|\mathcal{N}_g\right| - 2 \right) D \right] w_n + \left( \bar{s}_n - D \right) \sum_{m \neq n} w_m \geq 0, \nonumber \\
		&        & 
		\label{eqn:SFE_LineCapacity1}
		\sum_{n \in \mathcal{N}_g} \left[ \left( \bm{b}_n + \bm{f} \right) \cdot w_n \right] \geq \bm{0}, \nonumber \\
		\label{eqn:SFE_LineCapacity2}
		&        & 
		\sum_{n \in \mathcal{N}_g} \left[ \left( \bm{b}_n - \bm{f} \right) \cdot w_n \right] \leq \bm{0}, \nonumber 
	\end{eqnarray}
	where $\bm{b}_n \in \mathbb{R}^{|\mathcal{E}|}$ is a vector of constants defined as follows
	\begin{eqnarray}
		\bm{b}_n \triangleq 
		\sum_{m\in\mathcal{N}_g} \left[\mathbf{A}_g\right]_{*m} D - \left[\mathbf{A}_g\right]_{*n} \left( \left| \mathcal{N}_g - 1 \right| \right) D +
		\mathbf{A}_\ell \cdot \bm{d}.
	\end{eqnarray}
	
	We prove the existence of SFE by \cite[Theorem~4.1]{FacchineiKanzow-AnnalOR2010}. Specifically, we need to prove certain properties of the objective function and the feasible set of \eqref{eqn:SFE-reformulation}.
	
	First, we denote the feasible set of bids, namely the bids that satisfy the constraints in \eqref{eqn:SFE-reformulation} for all $n$, by $\mathcal{W}$. Based on Assumption~\ref{assumption:strictly-feasibility}, the set $\mathcal{W}$ is nonempty. Since the constraints in \eqref{eqn:SFE-reformulation} are linear constraints, the set $\mathcal{W}$ is convex and compact. We write $\mathcal{W}_n(\bm{w}_{-n})$ as the set of feasible bids from supplier $n$ given the others' bids $\bm{w}_{-n}$, namely $\mathcal{W}_n(\bm{w}_{-n}) \triangleq \left\{ w_n : (w_n, \bm{w}_{-n}) \in \mathcal{W} \right\}$. We also define $\bar{\mathcal{W}}_n \triangleq \left\{ w_n : \exists \bm{w}_{-n}~s.t.~(w_n, \bm{w}_{-n}) \in \mathcal{W} \right\}$. Since $\bar{\mathcal{W}}_n$ is the projection of $\mathcal{W}$ on the $w_n$-axis, the set $\bar{\mathcal{W}}_n$ is nonempty, convex, and compact. In addition, for any bidding profile $\bm{w}$ with $w_n \in \bar{\mathcal{W}}$ for every $n$, the set $\mathcal{W}_n(\bm{w}_{-n})$ is nonempty, convex, and compact, $\mathcal{W}_n(\bm{w}_{-n}) \subset \bar{\mathcal{W}}_n$, and $\mathcal{W}_n$, as a point-to-set mapping, is upper and lower semicontinuous. Therefore, the condition (a) in \cite[Theorem~4.1]{FacchineiKanzow-AnnalOR2010} is satisfied.
	
	Second, we show that supplier $n$'s payoff $u_n(w_n, \bm{w}_{-n})$ is concave in $w_n$ on $\mathcal{W}_n(\bm{w}_{-n})$. 
	We first note that any feasible bidding profile $\bm{w} \in \mathcal{W}$ must have at least two strictly positive bids. Suppose that all the suppliers bid zero, namely $w_n=0, \forall n$. Then each supplier $n$ produces $D$, resulting in a total supply of $\left|\mathcal{N}_g\right| D$. This supply profile violates the constraint that the supply equals the demand. Suppose that supplier $n$ is the only supplier with a strictly positive bid. Then supplier $n$ needs to supply
	$D - \frac{w_n}{w_n/\left[\left(\left|\mathcal{N}_g-1\right|\right)D\right]} = - \left(\left|\mathcal{N}_g-2\right|\right)D < 0$,
	which violates the constraint that $s_n \geq \underline{s}_n \geq 0$.
	Hence, at least two suppliers submit strictly positive bids.
	
	Now we can look at the concavity of the payoff function.
	Plugging \eqref{eqn:market-clearing-price} and \eqref{eqn:supply-as-function-of-bids} into the payoff function \eqref{eqn:payoff-function}, we can rewrite the payoff function as
	\begin{eqnarray}\label{eqn:payoff-detailed-expression}
		&  & u_n( w_n, \bm{w}_{-n} ) \\
		&=& \frac{\sum_{m\in\mathcal{N}_g} w_m}{\left|\mathcal{N}_g\right| - 1} 
		- w_n - c_n\left( S_n\left[p(w_n, \bm{w}_{-n}), w_n \right] \right). \nonumber
	\end{eqnarray}
	Since $c_n(s_n)$ is convex and increasing in $s_n$, and $S_n\left[ p( w_n,\bm{w}_{-n} ), w_n \right]$ is strictly convex in $w_n$ when $\sum_{m \neq n} w_m > 0$, the payoff function $u_n( w_n, \bm{w}_{-n} )$ is strictly concave in $w_n$ when $\sum_{m \neq n} w_m > 0$. We have just argued that $\sum_{m \neq n} w_m > 0$ at any feasible bidding profile. Hence, the payoff function $u_n( w_n, \bm{w}_{-n} )$ is strictly concave in $w_n$. Therefore, the condition (b) in \cite[Theorem~4.1]{FacchineiKanzow-AnnalOR2010} is satisfied.
	
	According to \cite[Theorem~4.1]{FacchineiKanzow-AnnalOR2010}, there exists an SFE.
	
	\begin{figure*}[!t]
		\normalsize
		\setcounter{MYtempeqncnt}{\value{equation}}
		\setcounter{equation}{\value{equation}}
		\begin{eqnarray}
			\frac{d c_n(s_n)}{d s_n} \cdot 
			\left( 1 + 
			\frac{s_n}{\left(|\mathcal{N}_g|-2\right) D}
			\right)
			- p^{\texttt{MCM}}
			- \underline{\mu}_n^{\texttt{MCM}} 
			+ \bar{\mu}_n^{\texttt{MCM}}
			- \left(\underline{\bm{\lambda}}^{\texttt{MCM}}\right)^T \cdot \left[\mathbf{A}_g\right]_{*n} 
			+ \left(\bar{\bm{\lambda}}^{\texttt{MCM}}\right)^T \cdot \left[\mathbf{A}_g\right]_{*n} 
			= 0 \label{eqn:MCM-KKT-Derivative-begin}\\
			\sum_{n\in\mathcal{N}_g} s_n = D \\
			\bm{0} \leq \bm{\underline{\mu}}^{\texttt{MCM}} \perp \bm{s} \geq \bm{0} \\
			\bm{0} \leq \bm{\bar{\mu}}^{\texttt{MCM}} \perp \bm{s} \leq \bm{\bar{s}} \\
			\bm{0} \leq \underline{\bm{\lambda}}^{\texttt{MCM}} \perp \left(\mathbf{A}_g \cdot \bm{s} + \mathbf{A}_\ell \cdot \bm{d}  + \bm{f}\right) \geq \bm{0} \\
			\bm{0} \leq \bar{\bm{\lambda}}^{\texttt{MCM}} \perp \left(\mathbf{A}_g \cdot \bm{s} + \mathbf{A}_\ell \cdot \bm{d}  - \bm{f}\right) \leq \bm{0} \label{eqn:MCM-KKT-Derivative-end}
		\end{eqnarray}
		\setcounter{equation}{\value{equation}}
		\hrulefill
		\vspace*{4pt}
	\end{figure*}
	
	\subsection{Equivalence} 
	Next, we prove that any equilibrium supply profile is the unique solution to the modified cost minimization problem \eqref{eqn:MCM}.
	
	\emph{Step 1: Since the payoff function $u_n(w_n, \bm{w}_{-n})$ is strictly concave in $w_n$, and since the constraints are linear in $w_n$, each supplier $n$'s optimization problem is a convex optimization problem. Therefore, $\bm{w}$ is a SFE if and only if it satisfies the KKT conditions for all $n$.}
	
	We define the Lagrangian multipliers associated with the four constraints in \eqref{eqn:SFE-reformulation} as $\underline{\mu}_n^{\texttt{SFE}} \in \mathbb{R}_+$, 
	$\bar{\mu}_n^{\texttt{SFE}} \in \mathbb{R}_+$,  
	$\bm{\underline{\lambda}}^{\texttt{SFE}} \in \mathbb{R}_+^{|\mathcal{E}|}$, 
	and $\bm{\bar{\lambda}}^{\texttt{SFE}} \in \mathbb{R}_+^{|\mathcal{E}|}$, respectively. 
	Then the action profile $\bm{w}$ is a SFE if and only if there exists $\underline{\mu}_n^{\texttt{SFE}}$, $\bar{\mu}_n^{\texttt{SFE}}$, $\bm{\underline{\lambda}}^{\texttt{SFE}}$, and $\bm{\bar{\lambda}}^{\texttt{SFE}}$ for all $n$ such that the KKT conditions in \eqref{eqn:SFE-KKT-Derivative-payoff}--\eqref{eqn:SFE-KKT-Derivative-line-capacity2} are satisfied for all $n$.
	
	Based on \eqref{eqn:supply-as-function-of-bids} and \eqref{eqn:payoff-detailed-expression}, we can compute the derivative of the payoff $u_n\left(w_n, \bm{w}_{-n}\right)$ as
	\begin{eqnarray}\label{eqn:derivative-of-payoff-function}
		&  & \frac{\partial u_n\left( w_n, \bm{w}_{-n} \right)}{\partial w_n} \\
		&=& -\frac{ |\mathcal{N}_g| - 2 }{ |\mathcal{N}_g| - 1 } + \frac{d c_n\left( S_n\left[p(\bm{w}), w_n \right] \right)}{d w_n} \nonumber \\
		&=& -\frac{ |\mathcal{N}_g| - 2 }{ |\mathcal{N}_g| - 1 } \nonumber \\
		&+& \left. \frac{d c_n\left( s_n \right)}{d s_n} \right|_{s_n = S_n\left[p(\bm{w}), w_n \right]} 
		\frac{\left( |\mathcal{N}_g|-1 \right) D \sum_{m\neq n} w_m}{\left( \sum_{m \in \mathcal{N}_g} w_m \right)^2}. \nonumber
	\end{eqnarray}
	
	Plugging \eqref{eqn:derivative-of-payoff-function} into \eqref{eqn:SFE-KKT-Derivative-payoff} and rearranging some terms, we get the following equality that is equivalent to \eqref{eqn:SFE-KKT-Derivative-payoff}:
	\begin{eqnarray}
		0 &=& 
		\left. \frac{d c_n\left( s_n \right)}{d s_n} \right|_{s_n = S_n\left[p(\bm{w}), w_n \right]}  \cdot
		\left( 1+ \frac{S_n\left[p(\bm{w}), w_n\right]}{\left( \left|\mathcal{N}_g\right| - 2 \right) D} \right) \nonumber \\
		&-& 
		p(\bm{w}) \nonumber \\
		&-&
		\left( 1 + \frac{ \underline{s}_n }{ \left(|\mathcal{N}_g|-2\right) D } \right) 
		\cdot \sum_{m\in\mathcal{N}_g} w_m
		\cdot \underline{\mu}_n^{\texttt{SFE}} \nonumber \\
		&+&
		\left( 1 + \frac{ \bar{s}_n }{ \left(|\mathcal{N}_g|-2\right) D } \right) 
		\cdot \sum_{m\in\mathcal{N}_g} w_m
		\cdot \bar{\mu}_n^{\texttt{SFE}} \nonumber \\
		&+& 
		\frac{ \sum_{m\in\mathcal{N}_g} w_m }{ \left(|\mathcal{N}_g|-2\right) D }
		\cdot \left(\bm{\underline{\lambda}}^{\texttt{SFE}}\right)^T \cdot (\bm{b}_n+\bm{f}) \nonumber \\
		&-&
		\frac{ \sum_{m\in\mathcal{N}_g} w_m }{ \left(|\mathcal{N}_g|-2\right) D }
		\cdot \left(\bm{\bar{\lambda}}^{\texttt{SFE}}\right)^T \cdot (\bm{b}_n-\bm{f}).
	\end{eqnarray}

	\emph{Step 2: Since the optimization problem $\texttt{(MCM)}$ is a convex optimization problem, $\bm{s}$ is a solution if and only if it satisfies the KKT conditions.}
	
	We write the Lagrangian multipliers associated with the constraints $\bm{s} \geq \bm{\underline{s}}$ and $\bm{s} \leq \bm{\bar{s}}$ in \eqref{eqn:MCM} as $\bm{\underline{\mu}}^{\texttt{MCM}}$ and $\bm{\bar{\mu}}^{\texttt{MCM}}$, respectively. Similarly, we write the Lagrangian multiplier associated with the constraint $\sum_{n \in \mathcal{N}_g} s_n = D$ in \eqref{eqn:MCM} as $p^{\texttt{MCM}}$. Finally, we write the Lagrangian multipliers associated with the constraints $-\bm{f} \leq \mathbf{A}_g \cdot \bm{s} + \mathbf{A}_\ell \cdot \bm{d} $ and $\mathbf{A}_g \cdot \bm{s} + \mathbf{A}_\ell \cdot \bm{d} \leq \bm{f}$ in \eqref{eqn:MCM} as $\underline{\bm{\lambda}}^{\texttt{MCM}}$ and $\bar{\bm{\lambda}}^{\texttt{MCM}}$, respectively.
	
	The KKT conditions are written in \eqref{eqn:MCM-KKT-Derivative-begin}--\eqref{eqn:MCM-KKT-Derivative-end}.

	\emph{Step 3: Given any SFE $\bm{w}$, the corresponding supply profile $\bm{s}$ with $s_n = S_n\left[ p(\bm{w}) , w_n \right]$ is a solution to the optimization problem \eqref{eqn:MCM}.}
	
	For any SFE $\bm{w}$, there exists $\underline{\mu}_n^{\texttt{SFE}}$, $\bar{\mu}_n^{\texttt{SFE}}$, $\bm{\underline{\lambda}}^{\texttt{SFE}}$, and $\bm{\bar{\lambda}}^{\texttt{SFE}}$ such that the KKT conditions in \eqref{eqn:SFE-KKT-Derivative-payoff}--\eqref{eqn:SFE-KKT-Derivative-line-capacity2} are satisfied.
	
	Define $s_n = S_n\left[ p(\bm{w}) , w_n \right]$ and the Lagrangian multipliers of the problem \texttt{(MCM)} as follows
	\begin{eqnarray}
		p^{\texttt{MCM}}  &=& p(\bm{w}) \\
		&\cdot&
		\left\{ 1 - 
		\frac{ \sum_{m} w_m }{ \left(|\mathcal{N}_g|-2\right) D }
		\left(\bm{\underline{\lambda}}^{\texttt{SFE}}\right)^T 
		\cdot 
		\left( \mathbf{A}_g D + \mathbf{A}_\ell \bm{d} + \bm{f} \right) 
		\right. \nonumber \\
		& &~+
		\left.
		\frac{ \sum_{m} w_m }{ \left(|\mathcal{N}_g|-2\right) D }
		\left(\bm{\bar{\lambda}}^{\texttt{SFE}}\right)^T 
		\cdot 
		\left( \mathbf{A}_g D + \mathbf{A}_\ell \bm{d} - \bm{f}\right) 
		\right\},
		\nonumber \\
		\underline{\mu}_n^{\texttt{MCM}} 
		&=& 
		\left( 1 + \frac{ \underline{s}_n }{ \left(|\mathcal{N}_g|-2\right) D } \right) 
		\cdot \sum_{m\in\mathcal{N}_g} w_m 
		\cdot \underline{\mu}_n^{\texttt{SFE}}, \\
		\bar{\mu}_n^{\texttt{MCM}}
		&=& 
		\left( 1 + \frac{ \bar{s}_n }{ \left(|\mathcal{N}_g|-2\right) D } \right) 
		\cdot \sum_{m\in\mathcal{N}_g} w_m
		\cdot \bar{\mu}_n^{\texttt{SFE}}, \\
		\underline{\bm{\lambda}}^{\texttt{MCM}}
		&=&
		\frac{ |\mathcal{N}_g|-1 }{ |\mathcal{N}_g|-2 } 
		\cdot \sum_{m\in\mathcal{N}_g} w_m
		\cdot
		\bm{\underline{\lambda}}^{\texttt{SFE}}, \\
		\bar{\bm{\lambda}}^{\texttt{MCM}}
		&=&
		\frac{ |\mathcal{N}_g|-1 }{ |\mathcal{N}_g|-2 } 
		\cdot \sum_{m\in\mathcal{N}_g} w_m
		\cdot \bm{\bar{\lambda}}^{\texttt{SFE}}.
	\end{eqnarray}
	
	Since $\bm{w}$, $p(\bm{w})$, $p^{\texttt{SFE}}$, $\underline{\mu}_n^{\texttt{SFE}}$, $\bar{\mu}_n^{\texttt{SFE}}$, $\underline{\bm{\lambda}}^{\texttt{SFE}}$, and $\bar{\bm{\lambda}}^{\texttt{SFE}}$ satisfy the KKT conditions in \eqref{eqn:MCM-KKT-Derivative-begin}--\eqref{eqn:MCM-KKT-Derivative-end}, it is not difficult to see that $\bm{s}$, $p^{\texttt{MCM}}$, $\underline{\mu}_n^{\texttt{MCM}}$, $\bar{\mu}_n^{\texttt{MCM}}$, $\underline{\bm{\lambda}}^{\texttt{MCM}}$, and $\bar{\bm{\lambda}}^{\texttt{MCM}}$ defined above satisfy the KKT conditions of \texttt{(MCM)} in \eqref{eqn:MCM-KKT-Derivative-begin}--\eqref{eqn:MCM-KKT-Derivative-end}. Therefore, the supply profile $\bm{s}$ is the solution to the problem \texttt{(MCM)}. This concludes our proof.
	
	\section{Proof of Theorem~\ref{theorem:PoA}}\label{proof:PoA}
	
	\subsection{Preparation}
	First, we show that for any $s_n \in \left[ \underline{s}_n, \bar{s}_n \right]$,
	\begin{eqnarray}\label{eqn:relationship_cost_modifiedcost}
		c_n(s_n) \leq \hat{c}_n(s_n) \leq \left( 1 + \frac{s_n}{\left(|\mathcal{N}_g| - 2\right) D} \right) \cdot c_n(s_n).
	\end{eqnarray}
	
	For convenience, we rewrite $\hat{c}_n(s_n)$ here:
	\begin{eqnarray*}
		\hat{c}_n(s_n) = \left( 1 + \frac{s_n}{\left(|\mathcal{N}_g| - 2\right) D} \right) \cdot c_n(s_n) 
		- \frac{\int_0^{s_n} c_n(x) dx}{\left(|\mathcal{N}_g| - 2\right) D}.
	\end{eqnarray*}
	
	It is clear that $\hat{c}_n(s_n) \leq \left( 1 + \frac{s_n}{\left(|\mathcal{N}_g| - 2\right) D} \right) \cdot c_n(s_n)$ because the integral in the expression of $\hat{c}_n(s_n)$ is nonnegative.
	
	To see $c_n(s_n) \leq \hat{c}_n(s_n)$, we look at
	$$
	\hat{c}_n(s_n) - c_n(s_n) = \frac{s_n}{\left(|\mathcal{N}_g| - 2\right) D}\cdot c_n(s_n) 
	- \frac{\int_0^{s_n} c_n(x) dx}{\left(|\mathcal{N}_g| - 2\right) D}.
	$$
	Note that $\hat{c}_n(0) - c_n(0) = 0$, and that the derivative of $\hat{c}_n(s_n) - c_n(s_n)$ with respect to $s_n$ is
	$$
	\frac{s_n}{\left(|\mathcal{N}_g| - 2\right) D}\cdot \frac{d c_n(s_n)}{d s_n} \geq 0.
	$$
	Therefore, we have $\hat{c}_n(s_n) - c_n(s_n) \geq 0$ for all $s_n$.
	
	Denote the socially optimal supply profile by $\bm{s}^*$ and the equilibrium supply profile by $\bm{s}^{**}$. Using \eqref{eqn:relationship_cost_modifiedcost} and the property of the equilibrium supply profile, we have the following inequality
	\begin{eqnarray}
		\sum_{n \in \mathcal{N}_g} c_n(s_n^{**}) 
		& \leq & \sum_{n \in \mathcal{N}_g} \hat{c}_n(s_n^{**}) \nonumber \\
		& \leq & \sum_{n \in \mathcal{N}_g} \hat{c}_n(s_n^{*}) \nonumber \\
		& \leq & \sum_{n \in \mathcal{N}_g} \left( 1 + \frac{s_n^*}{\left(|\mathcal{N}_g| - 2\right) D} \right) \cdot c_n(s_n^{*}) \nonumber \\
		& \leq & \max_{n \in \mathcal{N}_g} \left\{ 
		1 + \frac{s_n^*}{\left(|\mathcal{N}_g| - 2\right) D} 
		\right\}
		\cdot \sum_{n \in \mathcal{N}_g} c_n(s_n^{*}), \nonumber
	\end{eqnarray}
	where the first and third inequalities come from the inequality \eqref{eqn:relationship_cost_modifiedcost}, and the second inequality follows from the fact that $(s_n^{**})_{n \in \mathcal{N}_g}$ is the optimal solution to the problem \texttt{(MCM)}.
	
	Based on the above inequality, we can see that an upper bound of the PoA is 
	\begin{eqnarray}\label{eqn:PoA_UpperBound_Initial}
		1 + 
		\frac{\max_{n \in \mathcal{N}_g} s_n^*}
		{\left(|\mathcal{N}_g| - 2\right) D}.
	\end{eqnarray}
	Our next step is to compute this upper bound analytically.
	
	\subsection{Upper Bounds of PoA}
	To calculate the upper bound of PoA in \eqref{eqn:PoA_UpperBound_Initial} analytically, we need to calculate the socially optimal supply profile analytically, which is a difficult task. Therefore, we aim to further relax the upper bound in \eqref{eqn:PoA_UpperBound_Initial}. Instead of using the socially optimal supply profile $\bm{s}^*$ in \eqref{eqn:PoA_UpperBound_Initial}, we replace $\bm{s}^*$ with an arbitrary feasible supply profile $\bm{s}$, and find the maximum of the resulting term over all feasible supply profiles $\bm{s}$, namely:
	\begin{eqnarray}\label{eqn:upper-bound-feasible-supply-initial}
		& \displaystyle\max_{\bm{s}}  & 1 + \frac{\max_{n \in \mathcal{N}_g} s_n}{\left(|\mathcal{N}_g| - 2\right) D} \\
		& s.t.                    & \textstyle\sum_{n \in \mathcal{N}_g} s_n = D, \nonumber \\
		&                          & \underline{s}_n \leq s_n \leq \bar{s}_n, \quad\forall n \in \mathcal{N}_g, \nonumber \\
		&                          & -\bm{f} \leq \mathbf{A}_g \cdot \bm{s} + \mathbf{A}_\ell \cdot \bm{d} \leq \bm{f}. \nonumber
	\end{eqnarray}
	The optimal value of the above optimization problem must be an upper bound of \eqref{eqn:PoA_UpperBound_Initial}, because the socially optimal supply profile $\bm{s}^*$ is a feasible solution to the problem \eqref{eqn:upper-bound-feasible-supply-initial}.
	
	The optimization problem \eqref{eqn:upper-bound-feasible-supply-initial} can be decomposed into $\left| \mathcal{N}_g \right|$ problems, with the $n$th problem looking for the maximum of $1 + \frac{s_n}{\left(|\mathcal{N}_g| - 2\right) D}$
	over all feasible supply profiles, which is equivalent to looking for the maximum feasible supply $s_n$ from supplier $n$. Therefore, we will focus on finding supplier $n$'s maximum feasible supply, which is formulated as follows:
	\begin{eqnarray}\label{eqn:upper-bound-feasible-supply}
		& \displaystyle\max_{\bm{s}} & s_n \\
		& s.t.                   & \textstyle\sum_{m \in \mathcal{N}_g} s_m = D, \nonumber \\
		&                         & \underline{s}_m \leq s_m \leq \bar{s}_m, \quad\forall m \in \mathcal{N}_g, \nonumber \\
		&                         & -\bm{f} \leq \mathbf{A}_g \cdot \bm{s} + \mathbf{A}_\ell \cdot \bm{d} \leq \bm{f}. \nonumber
	\end{eqnarray}
	
	The first two constraints in \eqref{eqn:upper-bound-feasible-supply} imply that
	$$
	\underline{s}_n \leq s_n \leq \min\left\{ \bar{s}_n , D - \textstyle\sum_{m \neq n} \underline{s}_m \right\}.
	$$
	We replace the first two constraints in \eqref{eqn:upper-bound-feasible-supply} with the above inequality, which is a relaxation of the two original constraints. Then the optimal value of \eqref{eqn:upper-bound-feasible-supply} is upper bounded by the optimal value of the following optimization problem:
	\begin{eqnarray}\label{eqn:upper-bound-feasible-supply-relax}
		& \displaystyle\max_{\bm{s}} & s_n \\
		& s.t.                   & \underline{s}_n \leq s_n \leq \min\left\{ \bar{s}_n , D - \textstyle\sum_{m \neq n} \underline{s}_m \right\}, \nonumber \\
		&                         & -\bm{f} \leq \mathbf{A}_g \cdot \bm{s} + \mathbf{A}_\ell \cdot \bm{d} \leq \bm{f}. \nonumber
	\end{eqnarray}

	Since the shift matrix $\mathbf{A}$ is complicated and does not reflect the network topology, it is hard to solve \eqref{eqn:upper-bound-feasible-supply-relax} directly. To derive an equivalent formulation of \eqref{eqn:upper-bound-feasible-supply-relax}, we introduce new variables $\bm{p} = \left( p_{ij} \right)_{(i,j) \in \mathcal{E}}$ to denote the flows on each line, where $p_{ij}$ is the flow from bus $i$ to bus $j$. In addition, we write $\mathcal{E}(C)$ as the set of edges in a cycle $C$. Then we have an equivalent formulation of \eqref{eqn:upper-bound-feasible-supply-relax} as follows.
	\begin{eqnarray}\label{eqn:upper-bound-feasible-supply-relax-flow-variables}
		& \displaystyle\max_{\bm{s}, \bm{p}} & s_n \\
		& s.t.                               & \underline{s}_n \leq s_n \leq \min\left\{ \bar{s}_n , 
		D - \textstyle\sum_{m \neq n} \underline{s}_m 
		\right\}, \nonumber \\
		&                                     & -f_{ij} \leq p_{ij} \leq f_{ij},~\forall (i,j) \in \mathcal{E}, \nonumber \\
		&                                     & \sum_{(i,j) \in \mathcal{E}(C)} \frac{p_{ij}}{B_{ij}} = 0,~\mathrm{for~any~cycle}~C, \nonumber \\
		&                                     &  s_m - d_m = \textstyle\sum_{i \in \mathcal{N}(m)} p_{mi}, \forall m \in \mathcal{N}_g. \nonumber
	\end{eqnarray}
	Note that the second constraint in \eqref{eqn:upper-bound-feasible-supply-relax} is replaced by three new constraints, namely flow limit constraints (the second constraint), voltage angle constraints (the third one), and energy conservation at each node (the fourth one). 
	
	Since the objective function of \eqref{eqn:upper-bound-feasible-supply-relax-flow-variables} depends on $s_n$ only, we remove the voltage angle constraints for the cycles that do not include supplier $n$, and the energy conservation constraints for suppliers other than $n$. The resulting optimization problem is
	\begin{eqnarray}\label{eqn:upper-bound-feasible-supply-relax-final}
		& \displaystyle\max_{\bm{s}, \bm{p}} & s_n \\
		& s.t.                               & \underline{s}_n \leq s_n \leq \min\left\{ \bar{s}_n , D - \textstyle\sum_{m \neq n} \underline{s}_m \right\}, \nonumber \\
		&                                     & -f_{ij} \leq p_{ij} \leq f_{ij},~\forall (i,j) \in \mathcal{E}, \nonumber \\
		&                                     & \sum_{(i,j) \in \mathcal{E}(C)} \frac{p_{ij}}{B_{ij}} = 0,~\forall~\mathrm{cycle}~C~\mathrm{such~that}~n \in C, \nonumber \\
		&                                     &  s_n - d_n = \textstyle\sum_{i \in \mathcal{N}(n)} p_{ni}. \nonumber
	\end{eqnarray}
	
	It is not difficult to see that the optimal value of \eqref{eqn:upper-bound-feasible-supply-relax-final} is 
	\begin{eqnarray}\label{eqn:upper-bound-feasible-supply-relax-final-optval}
		\min\left\{ \bar{s}_n , D - \textstyle\sum_{m \neq n} \underline{s}_m , \hat{s}_n \right\},
	\end{eqnarray}
	where $\hat{s}_n$ is the optimal value of the following problem:
	\begin{eqnarray}\label{eqn:upper-bound-feasible-supply-relax-final-decomposed}
		& \displaystyle\max_{\bm{s}, \bm{p}} & s_n \\
		& s.t.                               & -f_{ij} \leq p_{ij} \leq f_{ij},~\forall (i,j) \in \mathcal{E}, \nonumber \\
		&                                     & \sum_{(i,j) \in \mathcal{E}(C)} \frac{p_{ij}}{B_{ij}} = 0,~\forall~\mathrm{cycle}~C~\mathrm{such~that}~n \in C, \nonumber \\
		&                                     & s_n - d_n = \textstyle\sum_{i \in \mathcal{N}(n)} p_{ni}. \nonumber
	\end{eqnarray}
	
	Now it remains to calculate $\hat{s}_i$ or its upper bound. First, the optimization problem \eqref{eqn:upper-bound-feasible-supply-relax-final-decomposed} is equivalent to the one below 
	\begin{eqnarray}\label{eqn:upper-bound-feasible-supply-relax-final-decomposed-equivalent}
		& \displaystyle\max_{\bm{p}} & d_n + \sum_{i \in \mathcal{N}(n)} p_{ni} \\
		& s.t.                               & -f_{ij} \leq p_{ij} \leq f_{ij},~\forall (i,j) \in \mathcal{E}, \nonumber \\
		&                                     & \sum_{(i,j) \in \mathcal{E}(C)} \frac{p_{ij}}{B_{ij}} = 0,~\forall~\mathrm{cycle}~C~\mathrm{such~that}~n \in C. \nonumber
	\end{eqnarray}
	
	We relax \eqref{eqn:upper-bound-feasible-supply-relax-final-decomposed-equivalent} by imposing the voltage angle constraints on some, but not all, cycles that include node $n$. In Definition~\ref{definition:partition}, we defined a partition $\mathcal{P}(n)$ of node $n$'s neighbors, where the neighbors are partitioned into subsets of singletons and duples. For each duple $\{i,j\} \in \mathcal{P}(n)$, we defined an associated cycle $\mathcal{C}_n(\{i,j\})$. To relax \eqref{eqn:upper-bound-feasible-supply-relax-final-decomposed-equivalent}, we will only impose the voltage angle constraints on these cycles. Specifically, the relaxed problem is written as follows:
	\begin{eqnarray}\label{eqn:upper-bound-feasible-supply-relax-final-decomposed-relax}
		& \displaystyle\max_{\bm{p}} & d_n + \sum_{i \in \mathcal{N}(n)} p_{ni} \\
		& s.t.                               & -f_{ij} \leq p_{ij} \leq f_{ij},~\forall (i,j) \in \mathcal{E}, \nonumber \\
		&                                     & \sum_{(m,\ell) \in \mathcal{E}\left[\mathcal{C}_n(\{i,j\})\right]} \frac{p_{m,\ell}}{B_{m,\ell}} = 0,~\forall~\{i,j\} \in \mathcal{P}(n). \nonumber
	\end{eqnarray}
	
	Our definition of the partition makes sure that any cycles $\mathcal{C}_n(\{i,j\})$ and $\mathcal{C}_n(\{m,\ell\})$ with $\{i,j\} \neq \{m,\ell\}$ do not share an edge. Therefore, the problem \eqref{eqn:upper-bound-feasible-supply-relax-final-decomposed-relax} can be decomposed into $|\mathcal{P}(n)|$ subproblems, each of which is determined by an element in the partition $\mathcal{P}(n)$. For any singleton set $\{i\} \in \mathcal{P}(n)$, the subproblem is
	\begin{eqnarray}\label{eqn:upper-bound-feasible-supply-relax-final-decomposed-relax-subproblem1}
		& \displaystyle\max_{p_{ni}} & p_{ni} \\
		& s.t.                                     & -f_{ni} \leq p_{ni} \leq f_{ni}, \nonumber
	\end{eqnarray}
	whose solution is clearly $p_{ni}^* = f_{ni} = \hat{f}_{ni}$, where $\hat{f}_{ni}$ is the effective flow limit from node $n$ to node $i$.
	For any duple set $\{i,j\} \in \mathcal{P}(n)$, the subproblem is
	\begin{eqnarray}\label{eqn:upper-bound-feasible-supply-relax-final-decomposed-relax-subproblem2}
		& \displaystyle\max_{p_{ni}, p_{nj}} & p_{ni} + p_{nj} \\
		& s.t.                                                & -f_{ni} \leq p_{ni} \leq f_{ni}, \nonumber \\
		&                                                      & -f_{nj} \leq p_{nj} \leq f_{nj}, \nonumber \\
		&                                                      & \sum_{(m,\ell) \in \mathcal{E}\left[\mathcal{C}_n(\{i,j\})\right]} \frac{p_{m,\ell}}{B_{m,\ell}} = 0, \nonumber
	\end{eqnarray}
	whose optimal value is upper bounded by $\hat{f}_{ni} + \hat{f}_{nj}$.
	
	In summary, an upper bound of the optimal value of \eqref{eqn:upper-bound-feasible-supply-relax-final-decomposed-relax} is $d_n + \sum_{m \in \mathcal{N}(n)} \hat{f}_{nm}$, which is also an upper bound of the optimal value of \eqref{eqn:upper-bound-feasible-supply-relax-final-decomposed-equivalent} and that of its equivalent problem \eqref{eqn:upper-bound-feasible-supply-relax-final-decomposed}.
	Accdoring to \eqref{eqn:upper-bound-feasible-supply-relax-final-optval}, an upper bound of the optimal value of \eqref{eqn:upper-bound-feasible-supply-relax-final} is then $\min\left\{ \bar{s}_n , D - \textstyle\sum_{m \neq n} \underline{s}_m , d_n + \sum_{m \in \mathcal{N}(n)} \hat{f}_{nm} \right\}$, which is also an upper bound of the optimal value of \eqref{eqn:upper-bound-feasible-supply-relax-flow-variables} and an upper bound of the optimal value of \eqref{eqn:upper-bound-feasible-supply}.
	Therefore, an upper bound of the optimal value of \eqref{eqn:upper-bound-feasible-supply-initial} is
	$$
	1 + \max_{n \in \mathcal{N}_g} \frac{ \min\left\{ \bar{s}_n , 
		D - \textstyle\sum_{m \neq n} \underline{s}_m , 
		d_n + \sum_{m \in \mathcal{N}(n)} \hat{f}_{nm} 
		\right\}
	}
	{ \left(|\mathcal{N}_g| - 2\right) D } ,
	$$
	which is an upper bound of the PoA. This concludes the proof of Theorem~\ref{theorem:PoA}.
	
	\section{Proof of Theorem~\ref{theorem:tightness}}\label{proof:tightness}
	
	\subsection{Preparation and Outline of The Proof}
	We consider an arbitrary weakly-cyclic network $(\mathcal{N}, \mathcal{E})$. We will find instances of demand profiles $\bm{d}$, sets of generators $\mathcal{N}_g$, cost functions $\{c_n\}_{n\in\mathcal{N}_g}$, generator capacity limits $\{\underline{s}_n, \bar{s}_n\}_{n\in\mathcal{N}_g}$, flow limits $\{f_{nm}\}_{nm \in \mathcal{E}}$, and admittances $\{B_{nm}\}_{nm \in \mathcal{E}}$, under which our upper bound of the PoA 
	can be made arbitrarily close to the actual PoA as we vary some of these parameters.
	
	We first fix the following parameters:
	\begin{eqnarray}\label{eqn:assumption_1}
	    \mathcal{N}_g = \mathcal{N} \triangleq N, ~\text{and}~ d_n = \frac{D}{N}, \bar{s}_n = D, \underline{s}_n = 0,~\forall n.
	\end{eqnarray}
	In words, there is a supplier on each node; the total demand is equally distributed to all the nodes;
	and the generation capacity constraints are never binding.
	Note that we can easily relax some of these restrictions (e.g., the upper limit of generation capacity $\bar{s}_n$ can be set lower). But we restrict to the simplified scenarios to make the proof easier to follow.
	
	For any graph $\mathcal{G}$, we can find its {\it spanning tree} $\mathcal{T}$, namely a tree that spans the graph (i.e., includes all vertices of $\mathcal{G}$) and is a subgraph of $\mathcal{G}$ \cite[pp.~55]{deo2017graph}. When there are multiple spanning trees, we can choose one arbitrarily. We then find the {\it centers} of the chosen spanning tree $\mathcal{T}$. The center of a graph is a node with minimal eccentricity, where the eccentricity of a node is defined as the greatest distance from this node to any other node \cite[pp.~46]{deo2017graph}. A tree may have either one center or two adjacent nodes as centers. In the case of two centers, we can pick either one of the two.
	
	Without loss of generality, we assume that node $1$ is the center of the spanning tree. We can always satisfy this assumption by changing the indices of the nodes.
	
	
	Next, we define cost functions. Supplier $1$ has the following cost function:
	\begin{eqnarray}
		c_1(s_1) = \left\{ \begin{array}{ll} 
			\delta \cdot s_1,          & s_1 \leq t \\
			s_1 - t + \delta \cdot t, & s_1 >    t
		\end{array}
		\right.,
	\end{eqnarray}
	which is piece-wise linear with two pieces. We set $\delta < 1$ so that the cost function is convex. All the other suppliers have the same cost function:
	\begin{eqnarray}
		c_n(s_n) = \alpha \cdot s_n, ~\forall n \geq 2,
	\end{eqnarray}
	where
	\begin{eqnarray}\label{eqn:definition_alpha}
		\alpha = \frac{ 1 + t / \left[ (N-2) D \right] }{ 1 + (D-t) / \left[ (N-1) (N-2) D \right] }.
	\end{eqnarray}
	We set $t > \frac{D}{N}$ so that $\alpha > 1$.
	As a result, supplier $1$, located at the center of the spanning tree of the network, has a lower generation cost than other suppliers at any supply level.
	
	At this point, the flow limits $\{f_{nm}\}_{nm \in \mathcal{E}}$ and admittances $\{B_{nm}\}_{nm \in \mathcal{E}}$ are the only parameters to be determined. We will show how to choose these parameters in Appendix~\ref{proof:tightness:supply_profile}, and show how to make the gap between the PoA and its upper bound arbitrarily small in Appendix~\ref{proof:tightness:PoA}.

	\subsection{Equilibrium and Socially Optimal Supply Profiles}\label{proof:tightness:supply_profile}
	
	In this part, we will find flow limits $\{f_{nm}\}_{nm \in \mathcal{E}}$ and admittances $\{B_{nm}\}_{nm \in \mathcal{E}}$, under which the equilibrium supply profile $\bm{s}^{**}$ is
	\begin{eqnarray}\label{eqn:equilibrium_supply}
		s_1^{**} = t, ~~ s_n^{**} = \frac{ D - s_1^{**} }{ N-1 }, ~\forall n \geq 2,
	\end{eqnarray}
	and the socially optimal supply profile $\bm{s}^*$ is
	\begin{eqnarray}\label{eqn:optimal_supply}
		s_1^* = \frac{D}{N} + \Delta, ~~ s_n^* = \frac{ D - s_1^* }{ N - 1 } = \frac{D}{N} - \frac{\Delta}{N-1}, ~\forall n \geq 2,
	\end{eqnarray}
	where $\Delta$ is chosen such that $s_1^{**} < s_1^* < D$.
	
	We will work on radial networks first, 
	and then extend to general weakly-cyclic networks. 
	
	\subsubsection{Radial Networks}
	For any radial network $\mathcal{G}$, the spanning tree $\mathcal{T}$ is itself. We define node $1$ (i.e., the center of the spanning tree) as the {\it root} of the tree, and denote the set of {\it descendants} of a node $i$ by $\mathcal{H}(i)$. Then for any edge $i,j$ with $j$ being the child of $i$, we assign the flow limit to be
	\begin{eqnarray}\label{eqn:flow_limit_radial}
		f_{ij} = \frac{\left|\mathcal{H}(j)\right|+1}{N-1} \cdot \Delta,~\text{with}~j \in \mathcal{H}(i).
	\end{eqnarray}
	Therefore, the flow limit of an edge is proportional to the number of nodes below the edge.
	
	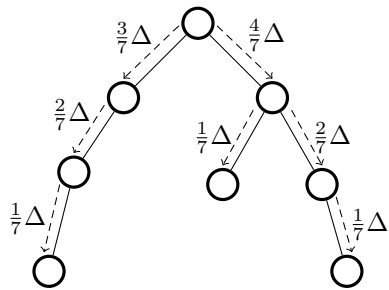
\begin{figure}
		\centering
		\begin{tikzpicture}[
				scale=0.33,
				root/.style={
					circle,
					minimum size=4mm,
					very thick,
					draw=black, 
					fill=white, 
					font=\large
				},
				branch1/.style={
					circle,
					minimum size=6mm,
					very thick,
					draw=red!50, 
					fill=red!20, 
					font=\large
				},
				branch2/.style={
					circle,
					minimum size=6mm,
					very thick,
					draw=blue!50, 
					fill=blue!20, 
					font=\large
				},
				branch3/.style={
					circle,
					minimum size=6mm,
					very thick,
					draw=black!50, 
					fill=black!20, 
					font=\large
				},
				output/.style={
					rectangle,minimum size=6mm,rounded corners=3mm,
					very thick,draw=black,
					font=\large},
				]

				\node (node1) at (0, 0) [root] {};
				\node (node2) at (-3, -3) [root] {};
				\node (node3) at (-5, -6) [root] {};
				\node (node4) at (-6, -10) [root] {};
				\node (node5) at (3, -3) [root] {};
				\node (node6) at (1, -6.5) [root] {};
				\node (node7) at (5, -6.5) [root] {};
				\node (node8) at (6, -10) [root] {};
				
				\draw [-] (node1) -- (node2);
				\draw [-] (node2) -- (node3);
				\draw [-] (node3) -- (node4);
				\draw [-] (node1) -- (node5);
				\draw [-] (node5) -- (node6);
				\draw [-] (node5) -- (node7);
				\draw [-] (node7) -- (node8);
				
				\draw[densely dashed, ->] 
				    ($(node1.west) + (-.1, -.1)$) -- 
				        node[above left=-0.2] {$\frac{3}{7} \Delta$} 
				    ($(node2.north) + (0, 0.1)$);
				
				\draw[densely dashed, ->] 
				    ($(node2.west) + (-.1, -.3)$) -- 
				        node[above left=-0.2] {$\frac{2}{7} \Delta$} 
				    ($(node3.north) + (0, 0.1)$);
				
				\draw[densely dashed, ->] 
				    ($(node3.west) + (.1, -.5)$) -- 
				        node[left=-0.05] {$\frac{1}{7} \Delta$} 
				    ($(node4.north) + (-0.2, 0.1)$);
				
				\draw[densely dashed, ->] 
				    ($(node1.east) + (.1, -.1)$) -- 
				        node[above right=-0.15] {$\frac{4}{7} \Delta$} 
				    ($(node5.north) + (0, 0.1)$);
				
				\draw[densely dashed, ->] 
				    ($(node5.west) + (-.1, -.5)$) -- 
				        node[left=-0.05] {$\frac{1}{7} \Delta$} 
				    ($(node6.north) + (0, 0.1)$);
				
				\draw[densely dashed, ->] 
				    ($(node5.east) + (.1, -.5)$) -- 
				        node[right=-0.05] {$\frac{2}{7} \Delta$} 
				    ($(node7.north) + (0, 0.1)$);
				    
				\draw[densely dashed, ->] 
				    ($(node7.east) + (-.1, -.5)$) -- 
				        node[right=-0.05] {$\frac{1}{7} \Delta$} 
				    ($(node8.north) + (0.2, 0.1)$);
				
			\end{tikzpicture}

		\caption{Illustration of power flows (indicated by the dashed lines) in a radial network. Under the supply profile in \eqref{eqn:optimal_supply}, the flow on any edge is proportional to the number of nodes descending from this edge.}
		\label{fig:flow-radial-network}
		
	\end{figure}
	
	Under the supply profile $\bm{s}^*$ defined in \eqref{eqn:optimal_supply}, each node $n \geq 2$ (i.e., all nodes other than the root) has supply exactly $\frac{\Delta}{N-1}$ less than the demand. Therefore, under the supply profile $\bm{s}^*$, the actual power flow $p_{ij}^*$ on each edge is equal to the flow limit $f_{ij}$ (see Fig.~\ref{fig:flow-radial-network} for illustration). Any decrease in the supply at any node $n \geq 2$ would violate the flow limit constraints. Since supplier $1$ has lower costs than other suppliers, the supply profile in \eqref{eqn:optimal_supply} has the highest possible supply from node $1$ without violating the flow limit constraints. Therefore, the supply profile $\bm{s}^*$ in \eqref{eqn:optimal_supply} is the socially optimal supply profile.
	
	Under the supply profile $\bm{s}^{**}$ defined in \eqref{eqn:equilibrium_supply}, we have $s_1^{**} < s_1^*$ and hence $s_n^{**} > s_n^*$ for $n \geq 2$. In fact, the net flow going into node $n \geq 2$, namely the demand minus the supply, under the supply profile $\bm{s}^{**}$ is a fixed fraction of the net flow under $\bm{s}^{*}$. The fixed fraction can be calculated as
	$$
	\frac{d_n - s_n^{**}}{d_n - s_n^{*}} = \frac{t - D/N}{\Delta} < 1.
	$$
	Therefore, the power flow $p_{ij}^{**}$ on each edge under the supply profile $\bm{s}^{**}$ is 
	$$
	p_{ij}^{**} = \frac{t - D/N}{\Delta} \cdot p_{ij}^* < f_{ij}^*.
	$$
	In other words, the supply profile $\bm{s}^{**}$ in \eqref{eqn:equilibrium_supply} respects the flow limit constraints.
	
	In addition, we can check that the supply profile $\bm{s}^{**}$ satisfies 
	\begin{eqnarray}\label{eqn:KKT_equilibrium_unconstrained}
		\left. \frac{\partial^{-} \hat{c}_1(s_1)}{\partial s_1} \right|_{s_1 = s_1^{**}} 
		< \left. \frac{\partial^{+} \hat{c}_1(s_1)}{\partial s_1} \right|_{s_1 = s_1^{**}}
		= \left. \frac{\partial \hat{c}_n(s_n)}{\partial s_n} \right|_{s_n = s_n^{**}} \!\!
	\end{eqnarray}
	for any $n \geq 2$, where $\hat{c}_n$ is the modified cost function in \eqref{eqn:MCM}, and $\frac{\partial^{-} \hat{c}_1(s_1)}{\partial s_1}$ and $\frac{\partial^{+} \hat{c}_1(s_1)}{\partial s_1}$ are the left and right derivatives. So the supply profile $\bm{s}^{**}$ is the optimal solution to the modified cost minimization problem \eqref{eqn:MCM}, and therefore is the equilibrium supply profile.
	
	In summary, under the flow limits defined in \eqref{eqn:flow_limit_radial}, the supply profiles defined in \eqref{eqn:equilibrium_supply} and in \eqref{eqn:optimal_supply} are the equilibrium and the socially optimal supply profiles, respectively. Note that for radial networks, we can set the admittances $\{B_{nm}\}_{nm \in \mathcal{E}}$ arbitrarily because they do not affect the power flow.

	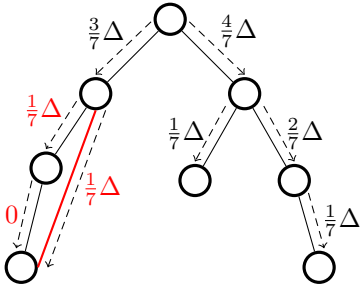
\begin{figure}
		\centering
			\begin{tikzpicture}[
				scale=0.33,
				root/.style={
					circle,
					minimum size=4mm,
					very thick,
					draw=black, 
					fill=white, 
					font=\large
				},
				branch1/.style={
					circle,
					minimum size=6mm,
					very thick,
					draw=red!50, 
					fill=red!20, 
					font=\large
				},
				branch2/.style={
					circle,
					minimum size=6mm,
					very thick,
					draw=blue!50, 
					fill=blue!20, 
					font=\large
				},
				branch3/.style={
					circle,
					minimum size=6mm,
					very thick,
					draw=black!50, 
					fill=black!20, 
					font=\large
				},
				output/.style={
					rectangle,minimum size=6mm,rounded corners=3mm,
					very thick,draw=black,
					font=\large},
				]

				\node (node1) at (0, 0) [root] {};
				\node (node2) at (-3, -3) [root] {};
				\node (node3) at (-5, -6) [root] {};
				\node (node4) at (-6, -10) [root] {};
				\node (node5) at (3, -3) [root] {};
				\node (node6) at (1, -6.5) [root] {};
				\node (node7) at (5, -6.5) [root] {};
				\node (node8) at (6, -10) [root] {};
				
				\draw [-] (node1) -- (node2);
				\draw [-] (node2) -- (node3);
				\draw [-] (node3) -- (node4);
				\draw [-] (node1) -- (node5);
				\draw [-] (node5) -- (node6);
				\draw [-] (node5) -- (node7);
				\draw [-] (node7) -- (node8);
				\draw [-, red, thick] (node2.south) -- (node4.east);
				
				\draw[densely dashed, ->] 
				    ($(node1.west) + (-.1, -.1)$) -- 
				        node[above left=-0.2] {$\frac{3}{7} \Delta$} 
				    ($(node2.north) + (0, 0.1)$);
				
				\draw[densely dashed, ->] 
				    ($(node2.west) + (-.1, -.3)$) -- 
				        node[above left=-0.2, red] {$\frac{1}{7} \Delta$} 
				    ($(node3.north) + (0, 0.1)$);
				
				\draw[densely dashed, ->] 
				    ($(node3.west) + (.1, -.5)$) -- 
				        node[left=-0.05, red] {$0$} 
				    ($(node4.north) + (-0.2, 0.1)$);
				
				\draw[densely dashed, ->] 
				    ($(node1.east) + (.1, -.1)$) -- 
				        node[above right=-0.15] {$\frac{4}{7} \Delta$} 
				    ($(node5.north) + (0, 0.1)$);
				
				\draw[densely dashed, ->] 
				    ($(node5.west) + (-.1, -.5)$) -- 
				        node[left=-0.05] {$\frac{1}{7} \Delta$} 
				    ($(node6.north) + (0, 0.1)$);
				
				\draw[densely dashed, ->] 
				    ($(node5.east) + (.1, -.5)$) -- 
				        node[right=-0.05] {$\frac{2}{7} \Delta$} 
				    ($(node7.north) + (0, 0.1)$);
				    
				\draw[densely dashed, ->] 
				    ($(node7.east) + (-.1, -.5)$) -- 
				        node[right=-0.05] {$\frac{1}{7} \Delta$} 
				    ($(node8.north) + (0.2, 0.1)$);
				
				\draw[densely dashed, ->] 
				    ($(node2.south) + (0.4, 0)$) -- 
				        node[right=-0.05, red] {$\frac{1}{7} \Delta$} 
				    ($(node4.east) + (0.4, 0)$);
				
			\end{tikzpicture}

		\caption{Illustration of power flows (indicated by the dashed lines) in a weakly-cyclic network, which is expanded from the radial network in Fig.~\ref{fig:flow-radial-network} by added one edge connecting two nodes in the same branch (marked in red). The figure shows the flows under the supply profile in \eqref{eqn:optimal_supply}, where the flows in the newly-formed cycle are adjusted from those in the radial network according to \eqref{eqn:flow_adjustment}--\eqref{eqn:flow_adjustment_calculation_special_case}. Note that the flows on the edges outside the cycle remain unchanged.}
		\label{fig:flow-weakly-cyclic-network-1}
		
	\end{figure}

	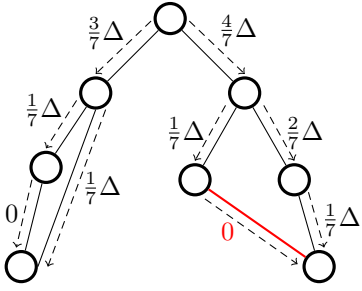
\begin{figure}
		\centering
			\begin{tikzpicture}[
				scale=0.33,
				root/.style={
					circle,
					minimum size=4mm,
					very thick,
					draw=black, 
					fill=white, 
					font=\large
				},
				branch1/.style={
					circle,
					minimum size=6mm,
					very thick,
					draw=red!50, 
					fill=red!20, 
					font=\large
				},
				branch2/.style={
					circle,
					minimum size=6mm,
					very thick,
					draw=blue!50, 
					fill=blue!20, 
					font=\large
				},
				branch3/.style={
					circle,
					minimum size=6mm,
					very thick,
					draw=black!50, 
					fill=black!20, 
					font=\large
				},
				output/.style={
					rectangle,minimum size=6mm,rounded corners=3mm,
					very thick,draw=black,
					font=\large},
				]

				\node (node1) at (0, 0) [root] {};
				\node (node2) at (-3, -3) [root] {};
				\node (node3) at (-5, -6) [root] {};
				\node (node4) at (-6, -10) [root] {};
				\node (node5) at (3, -3) [root] {};
				\node (node6) at (1, -6.5) [root] {};
				\node (node7) at (5, -6.5) [root] {};
				\node (node8) at (6, -10) [root] {};
				
				\draw [-] (node1) -- (node2);
				\draw [-] (node2) -- (node3);
				\draw [-] (node3) -- (node4);
				\draw [-] (node1) -- (node5);
				\draw [-] (node5) -- (node6);
				\draw [-] (node5) -- (node7);
				\draw [-] (node7) -- (node8);
				\draw [-] (node2.south) -- (node4.east);
				\draw [-, red, thick] (node6) -- (node8);
				
				\draw[densely dashed, ->] 
				    ($(node1.west) + (-.1, -.1)$) -- 
				        node[above left=-0.2] {$\frac{3}{7} \Delta$} 
				    ($(node2.north) + (0, 0.1)$);
				
				\draw[densely dashed, ->] 
				    ($(node2.west) + (-.1, -.3)$) -- 
				        node[above left=-0.2] {$\frac{1}{7} \Delta$} 
				    ($(node3.north) + (0, 0.1)$);
				
				\draw[densely dashed, ->] 
				    ($(node3.west) + (.1, -.5)$) -- 
				        node[left=-0.05] {$0$} 
				    ($(node4.north) + (-0.2, 0.1)$);
				
				\draw[densely dashed, ->] 
				    ($(node1.east) + (.1, -.1)$) -- 
				        node[above right=-0.15] {$\frac{4}{7} \Delta$} 
				    ($(node5.north) + (0, 0.1)$);
				
				\draw[densely dashed, ->] 
				    ($(node5.west) + (-.1, -.5)$) -- 
				        node[left=-0.05] {$\frac{1}{7} \Delta$} 
				    ($(node6.north) + (0, 0.1)$);
				
				\draw[densely dashed, ->] 
				    ($(node5.east) + (.1, -.5)$) -- 
				        node[right=-0.05] {$\frac{2}{7} \Delta$} 
				    ($(node7.north) + (0, 0.1)$);
				    
				\draw[densely dashed, ->] 
				    ($(node7.east) + (-.1, -.5)$) -- 
				        node[right=-0.05] {$\frac{1}{7} \Delta$} 
				    ($(node8.north) + (0.2, 0.1)$);
				
				\draw[densely dashed, ->] 
				    ($(node2.south) + (0.4, 0)$) -- 
				        node[right=-0.05] {$\frac{1}{7} \Delta$} 
				    ($(node4.east) + (0.4, 0)$);
				
				\draw[densely dashed, ->] 
				    ($(node6.south) + (0.4, -0.1)$) -- 
				        node[left=0.1, red] {$0$} 
				    ($(node8.west) + (-0.2, 0.1)$);
				
			\end{tikzpicture}

		\caption{Illustration of power flows (indicated by the dashed lines) in a weakly-cyclic network, which is expanded from the radial network in Fig.~\ref{fig:flow-weakly-cyclic-network-1} by added one edge connecting two nodes in different branches (marked in red). The figure shows the flows under the supply profile in \eqref{eqn:optimal_supply}, where the flows in the newly-formed cycle remain the same by setting the admittances according to \eqref{eqn:admittance_adjustment}. In this case, all the flows in the network remain the same.}
		\label{fig:flow-weakly-cyclic-network-2}
		
	\end{figure}

	\subsubsection{Weakly-Cyclic Networks}
	For a weakly-cyclic network $\mathcal{G}$, we first calculate the power flow on each edge under the supply profile $\bm{s}^{*}$ in \eqref{eqn:optimal_supply}. We will calculate the power flows on its spanning tree as baselines, and make adjustments to get the actual power flows on the weakly-cyclic network itself. 
	
	Based on the analysis for radial networks, if we injected the supply profile $\bm{s}^*$ into the spanning tree $\mathcal{T}$ of the network, the power flow on each edge would be 
	$$
	p_{ij}^{*\mathcal{T}} = \frac{\left|\mathcal{H}(j)\right|+1}{N-1} \cdot \Delta,~\text{with}~j \in \mathcal{H}(i).
	$$
	
	Each cycle in the weakly-cyclic network $\mathcal{G}$ can be obtained by adding one edge to its spanning tree. Since the cycles do not share any edge, we can analyze each cycle independently. There are two cases.
	
	\begin{itemize}
	    \item The new edge connects two nodes in the same branch (see Fig.~\ref{fig:flow-weakly-cyclic-network-1} for illustration): In this case, for any edge $(i,j)$ in the newly-added cycle $C$, we add an adjustment term $\omega_C^{*}$ to the flow on all edges in the cycle $C$, namely
	        \begin{eqnarray}\label{eqn:flow_adjustment}
	            p_{k \ell}^{*} = p_{k \ell}^{*\mathcal{T}} + \omega_C^{*},~\forall (k,\ell) \in \mathcal{E}(C),
	        \end{eqnarray}
	        where the adjustment term $\omega_C^{*}$ is chosen to satisfy the voltage angle constraint
	        \begin{eqnarray}\label{eqn:flow_adjustment_calculation}
	           \sum_{(k,\ell) \in \mathcal{E}(C)} \frac{p_{k\ell}^{*}}{B_{k\ell}} = 0.
	        \end{eqnarray}
	    Note that the adjustment $\omega_C^{*}$ can be calculated analytically based on $p_{k\ell}^{*\mathcal{T}}$ and $B_{k\ell}$. In the special case where all admittances are equal, we have 
	    \begin{eqnarray}\label{eqn:flow_adjustment_calculation_special_case}
	        \omega_C^{*} = - \left( \sum_{(k,\ell) \in \mathcal{E}(C)} p_{k\ell}^{*} \right) / \left|\mathcal{E}(C)\right|.
	    \end{eqnarray}
	    \item The new edge connects two nodes in different branches (see Fig.~\ref{fig:flow-weakly-cyclic-network-2} for illustration): In this case, we set the admittances so that the voltage angle constraint is satisfied by the flows in the spanning tree, namely
	    \begin{eqnarray}\label{eqn:admittance_adjustment}
	           \sum_{(k,\ell) \in \mathcal{E}(C)} \frac{p_{k\ell}^{*\mathcal{T}}}{B_{k\ell}} = 0.
	    \end{eqnarray}
	    There will always be solutions $B_{k\ell}$ to the above equation because the flows $p_{k\ell}$ on different branches have both positive and negative values. Since the baseline flows already satisfy the voltage angle constraint, the flow on the new edge is zero.
	\end{itemize}
	
	The power flow $p_{ij}^{*}$ on each edge of $\mathcal{G}$ can be calculated based on $p_{ij}^{*\mathcal{T}}$. If an edge $(i,j)$ does not belong to any cycle in $\mathcal{G}$, the flow is the same as that in the spanning tree, namely $p_{ij}^{*} = p_{ij}^{*\mathcal{T}}$. In addition, since the graph is weakly-cyclic, no edge belongs to more than one cycle. Therefore, we can deal with each cycle independently and make the adjustment only once for each edge.
	
	We will set the flow limits as
	\begin{eqnarray}\label{eqn:flow_limit_cyclic}
		f_{ij} = \left| p_{ij}^{*} \right|
	\end{eqnarray}
	for all edges. Note that we take the absolute value because due to the adjustment $\omega_C^{*}$, $p_{ij}^{*}$ may be negative even if $p_{ij}^{*\mathcal{T}}$ is positive. The flow limits set in \eqref{eqn:flow_limit_cyclic} ensure that the supply profile $\bm{s}^{*}$ respects the flow limit constraints.
	
	When the supply profile is $\bm{s}^{**}$ in \eqref{eqn:equilibrium_supply}, the power flow on each edge is still a fixed fraction of the flow under $\bm{s}^{*}$, namely
	$$
	p_{ij}^{**} = \frac{t - D/N}{\Delta} \cdot p_{ij}^*.
	$$
	Therefore, we have $\left| p_{ij}^{**} \right| < \left| p_{ij}^{*} \right| = f_{ij}$. So the supply profile $\bm{s}^{**}$ also respects the flow limit constraints.
	
	As in the case of radial networks, under the flow limits defined in \eqref{eqn:flow_limit_cyclic}, the supply profiles defined in \eqref{eqn:equilibrium_supply} and in \eqref{eqn:optimal_supply} are the equilibrium and the socially optimal supply profiles, respectively.

	\subsection{Tightness of The Upper Bound}\label{proof:tightness:PoA}
	
	Given the equilibrium supply profile $\bm{s}^{**}$ in \eqref{eqn:equilibrium_supply} and the socially optimal supply profile $\bm{s}^{*}$ in \eqref{eqn:optimal_supply}, we can calculate the PoA analytically as
	\begin{eqnarray}\label{eqn:PoA_analytical}
		\mathrm{PoA}(\delta) = \frac{ \delta \cdot t + \alpha \cdot (D-t) }
		{ \hat{d}_1 - t + \delta \cdot t + \alpha \cdot \left( D - \hat{d}_1 \right) },
	\end{eqnarray}
	where we define 
	$\hat{d}_1 \triangleq \frac{D}{N} + \Delta$ for notational simplicity. We write the PoA as a function of $\delta$, the parameter in supplier 1's cost function, to highlight its dependence on $\delta$.
	
	The following lemma specifies our upper bound of PoA.
	\begin{lemma}\label{lemma:PoA_upper_bound}
		Under the flow limits defined in \eqref{eqn:flow_limit_cyclic}, our upper bound of PoA is
		$1 + \frac{\hat{d}_1}{(N-2)D}$.
	\end{lemma}
	\begin{IEEEproof}
    We first show that the flow limits in \eqref{eqn:flow_limit_cyclic} satisfy
    \begin{eqnarray}\label{eqn:root_largest_flow_limit}
    d_1 + \textstyle\sum_{m \in \mathcal{N}(1)} \hat{f}_{1m} \geq d_n + \textstyle\sum_{m \in \mathcal{N}(n)} \hat{f}_{nm},~\forall n \geq 2.
    \end{eqnarray}
    
    For a radial network, the effective flow limit $\hat{f}_{nm}$ is equal to the physical flow limit $f_{nm}$ because there is no cycle. Consider an arbitrary node $n$ other than the root. If node $m$ is node $n$'s child, namely $m \in \mathcal{N}(n) \cap \mathcal{H}(n)$, we have
    $$
    f_{nm} = \frac{\left|\mathcal{H}(m)\right|+1}{N-1} \cdot \Delta.
    $$
    If node $m$ is node $n$'s parent, namely $m \in \mathcal{N}(n)$ and $n \in \mathcal{H}(m)$, we have
    $$
    f_{nm} = \frac{1 + \left[ \sum_{k \in \mathcal{N}(n) \cap \mathcal{H}(n)} 
                                  \left|\mathcal{H}(k)\right|+1 \right]}
                  {N-1} 
             \cdot \Delta.
    $$
    Therefore, we have
    $$
    \sum_{m \in \mathcal{N}(n)} f_{nm} = 
        \frac{1 + 2 \cdot \left[ \sum_{k \in \mathcal{N}(n) \cap \mathcal{H}(n)} 
                                    \left|\mathcal{H}(k)\right|+1 \right]}
             {N-1} 
             \cdot \Delta.
    $$
    For the root of the radial network, namely node $1$, we have
    \begin{eqnarray}
    \sum_{m \in \mathcal{N}(1)} f_{1m} 
        &=& \sum_{m \in \mathcal{N}(1) \cap \mathcal{H}(1)} f_{1m} \nonumber \\
        &=& \frac{\sum_{m \in \mathcal{N}(1) \cap \mathcal{H}(1)} 
                \left|\mathcal{H}(m)\right|+1}
             {N-1} 
             \cdot \Delta \nonumber \\
        &=& \frac{N-1}{N-1} \cdot \Delta = \Delta.
    \end{eqnarray}
    Since we have chosen the center of the network to be the root, the network is ``balanced'', in the sense that no branch has more than $N/2$ nodes. Therefore, we have
    \begin{eqnarray}
    \sum_{m \in \mathcal{N}(n)} f_{nm} 
        &=& 
            \frac{1 + 2 \cdot \left[ \sum_{k \in \mathcal{N}(n) \cap \mathcal{H}(n)} 
                                        \left|\mathcal{H}(k)\right|+1 \right]}
                 {N-1} 
            \cdot \Delta \nonumber \\
        &\leq& \frac{1 + 2 \cdot \left[ N/2 - 1 \right]}
                    {N-1} 
               \cdot \Delta \nonumber \\
        &=& \frac{N-1}{N-1} \cdot \Delta = \Delta, ~\forall n \geq 2.
    \end{eqnarray}
    So the inequality \eqref{eqn:root_largest_flow_limit} holds for radial networks.
    
    For a weakly-cyclic network, following the same logic as in radial networks, the power flow on its spanning tree satisfies
    $$
    \sum_{m \in \mathcal{N}(n)} \left|p_{nm}^{*\mathcal{T}}\right| \leq \Delta, ~\forall n \geq 2.
    $$
    We need to show that the above inequality holds for the actual power flow $p_{nm}^{*}$ after the adjustment. Since the cycles do not share an edge, we can check each cycle independently. Each cycle can be obtained by adding one edge to the spanning tree. 
    Suppose that the cycle is $i_1, i_2, \ldots, i_k, i_1$, and that the edge added to the spanning tree is between $i_1$ and $i_k$. We discuss two scenarios, depending on whether $i_1$ and $i_k$ are in the same branch or not.
    
    If $i_1$ and $i_k$ are in the same branch, one is the parent of the other. Without loss of generality, suppose that $i_1$ is the parent of $i_k$, then $i_1$ must be the parent of all other nodes in the cycle. After the adjustment, the sum of $i_1$'s flow limits do not change, while the sum of $i_j$'s flow limits decreases for all $j=2,\ldots,k$. Note that if the root node $1$ belongs to any cycle, it must be the parent of the nodes in the cycle (i.e., it is the ``$i_1$ node‘’). In summary, after adjustment, the sum of flow limits for node $1$ do not change, while the sums of flow limits for all other nodes remain the same or decrease. Therefore, the inequality \eqref{eqn:root_largest_flow_limit} still holds.
    
    If $i_1$ and $i_k$ are not in the same branch, we adjusted the admittances based on \eqref{eqn:admittance_adjustment}, so that the flows do not change. Therefore, the inequality \eqref{eqn:root_largest_flow_limit} still holds.
    
	Now we can derive the simplified expression of the upper bound in Theorem~\ref{theorem:PoA}:
	$$
	1 + \max_{n \in \mathcal{N}} \frac{ \min\left\{ \bar{s}_n , 
		D - \textstyle\sum_{m \neq n} \underline{s}_m , 
		d_n + \sum_{m \in \mathcal{N}(n)} \hat{f}_{nm} 
		\right\}
	}
	{ \left(N - 2\right) D }.
	$$
	
	Based on the generation capacity limits set in \eqref{eqn:assumption_1}, the terms $\bar{s}_n$ and $D - \textstyle\sum_{m \neq n} \underline{s}_m$ satisfy
	$$
	\bar{s}_n = D - \textstyle\sum_{m \neq n} \underline{s}_m = D > \hat{d}_1 \geq d_n + \sum_{m \in \mathcal{N}(n)} \hat{f}_{nm},~\forall n \geq 2.
	$$
	Therefore, the upper bound is equal to the network-dependent component:
	$$
	1 + 
	\max_{n \in \mathcal{N}} 
	\frac{
		d_n + \textstyle\sum_{m \in \mathcal{N}(n)} \hat{f}_{nm}
	}
	{ \left(N - 2\right) D }.
	$$
	Due to \eqref{eqn:root_largest_flow_limit}, the upper bound is equal to $1 + \frac{\hat{d}_1}{(N-2)D}$.
	\end{IEEEproof}
	
	With the analytical expressions of the PoA in \eqref{eqn:PoA_analytical} and the upper bound in Lemma~\ref{lemma:PoA_upper_bound}, we show that they can be made arbitrarily close. Specifically, we have
	\begin{eqnarray}
		&    & 1 + \frac{ \hat{d}_1 }{ \left(N - 2\right) D }
		- \mathrm{PoA}(\delta) \\
		& = & \left[ 1 + \frac{ \hat{d}_1 }{ \left(N - 2\right) D }
		- \mathrm{PoA}(0)
		\right]
		+ \left[ \mathrm{PoA}(0) - \mathrm{PoA}(\delta) \right]. \nonumber
	\end{eqnarray}
	We look at these two terms separately.
	
	For the first term, we have
	\begin{eqnarray}
		&    & 1 + \frac{ \hat{d}_1 }
		{ \left(N - 2\right) D }
		- \mathrm{PoA}(0) \\
		& = & 1 + \frac{ \hat{d}_1 }
		{ \left(N - 2\right) D }
		- \frac{ \alpha \cdot (D-t) }
		{ \hat{d}_1 - t + \alpha \cdot \left( D - \hat{d}_1 \right) } \nonumber \\
		& = & \frac{ \hat{d}_1 }{ \left(N - 2\right) D }
		- \frac{ (\alpha-1) \cdot (\hat{d}_1-t) }
		{ \hat{d}_1 - t + \alpha \cdot \left( D - \hat{d}_1 \right) } \nonumber \\
		& = & \frac{ \hat{d}_1 }{ \left(N - 2\right) D }
		- \frac{ \alpha-1 }
		{ 1 + \alpha \cdot \frac{ D - \hat{d}_1 }{ \hat{d}_1 - t } }. \nonumber
	\end{eqnarray}
	
	We let $t \uparrow \hat{d}_1$, $\hat{d}_1 \uparrow D$, while $\frac{ D - \hat{d}_1 }{ \hat{d}_1 - t } \downarrow 0$. Then from the definition of $\alpha$ in \eqref{eqn:definition_alpha}, we have $\alpha \rightarrow 1+ \frac{1}{N-2}$. We can see that $1 + \frac{ \hat{d}_1 }{ \left(N - 2\right) D } - \mathrm{PoA}(0) \rightarrow 0$.
	
	For the second term, we have
	\begin{eqnarray}
		&    & \mathrm{PoA}(0)- \mathrm{PoA}(\delta) \\
		& = & 1 + \frac{ (\alpha-1) \cdot (\hat{d}_1-t) }
		{ \hat{d}_1 - t + \alpha \cdot \left( D - \hat{d}_1 \right) } \nonumber \\
		& - &   \left[ 1 + \frac{ (\alpha-1) \cdot (\hat{d}_1-t) }
		{ \hat{d}_1 - t +\delta \cdot t + \alpha \cdot \left( D - \hat{d}_1 \right) } 
		\right] \nonumber \\
		& = & \frac{ (\alpha-1) \cdot (\hat{d}_1-t) \cdot \delta t }
		{ \left[ \hat{d}_1 - t + \alpha \cdot \left( D - \hat{d}_1 \right) \right] 
			\left[ \hat{d}_1 - t + \alpha \cdot \left( D - \hat{d}_1 \right) + \delta t \right] }
		\nonumber \\
		& = & \frac{ (\alpha-1) \cdot \delta t }
		{ \left[ 1 + \alpha \cdot \frac{ D - \hat{d}_1 }{ \hat{d}_1 - t } \right] 
			\left[ 1 + \alpha \cdot \frac{ D - \hat{d}_1 }{ \hat{d}_1 - t } + \frac{\delta t}{ \hat{d}_1 - t } \right] }. \nonumber
	\end{eqnarray}
	
	Taking the same limits of $t \uparrow \hat{d}_1$ and $\hat{d}_1 \uparrow D$ while $\frac{ D - \hat{d}_1 }{ \hat{d}_1 - t } \downarrow 0$, we can see that the numerator approaches $\frac{\delta \hat{d}_1}{N-2}$, while the denominator approaches $\infty$ (since the term $\frac{\delta t}{ \hat{d}_1 - t } \uparrow \infty$). Therefore, we have $\mathrm{PoA}(0)- \mathrm{PoA}(\delta) \rightarrow 0$.
	
	In summary, we have $1 + \frac{ \hat{d}_1 }{ \left(N - 2\right) D } - \mathrm{PoA}(\delta) \rightarrow 0$. This concludes the proof.
	
	\bibliographystyle{IEEEtran}
	\bibliography{IEEEabrv, xyz_power_TPS}
	
\end{document}